\documentclass[aps,prb,preprint,superscriptaddress,amsmath,amssymb,showpacs]{revtex4-1}%

\renewcommand{\AA}{\mathcal{A}}
\newcommand{\BB}{\mathcal{B}}
\newcommand{\CC}{\mathcal{C}}
\newcommand{\LL}{\mathcal{L}}

\newcommand{\TT}{\mathcal{T}}
\newcommand{\DD}{\mathcal{D}}
\newcommand{\FF}{\mathcal{F}}
\newcommand{\dd}{\mathrm{d}}

\newcommand{\ii}{\text{i}}

\newcommand{\tr}[2]{\text{Tr}_{ #1 } \left\{ #2 \right\}}

\newcommand{\av}[1]{\langle #1 \rangle}
\newcommand{\avg}[2]{\langle #1 \rangle_{#2}}
\newcommand{\expo}[1]{\text{exp}\left( #1 \right)}

\newcommand{\tS}{\text{S}}
\newcommand{\tL}{\text{L}}
\newcommand{\tR}{\text{R}}
\newcommand{\tI}{\text{I}}

\newcommand{\tB}{\text{B}}
\newcommand{\tSB}{\text{SB}}

\newcommand{\tot}{\text{tot}}

\newcommand{\vib}{\text{vib}}

\newcommand{\e}{\text{e}}
\newcommand{\el}{\text{el}}

\newcommand{\unit}[1]{\,\mathrm{#1}}

\newcommand{\gv}[1]{\ensuremath{\mbox{\boldmath$ #1 $}}}
\newcommand{\leads}{\text{leads}}
\newcommand{\pd}[2]{\frac{\partial #1}{\partial #2}}  
  
\usepackage{braket}
\usepackage{subfigure}
\usepackage{graphicx}
\usepackage{xcolor}
\usepackage{dsfont}
\usepackage{pstricks}

\graphicspath{{./overview_5e-2g_06lo/}{../../../HEOM/programs/el_vib_bath/adiabatic_regime/Plots/}{../../../HEOM/programs/system_vib_pola_trafo/Rainer_PHD_model/Plots/}{../../../HEOM/programs/Pade_decomposition/Results/}{./Plots_mm_F_V/}{./Plots_1m_fig/}{./Plots_BDBT/}}

\begin{document}



\title{Hierarchical quantum master equation approach to electronic-vibrational coupling in nonequilibrium transport through nanosystems: Reservoir formulation and application to vibrational instabilities}

\author{C.\ Schinabeck}
\affiliation{Institut f\"ur Theoretische Physik und Interdisziplin\"ares Zentrum f\"ur Molekulare
Materialien, \\
Friedrich-Alexander-Universit\"at Erlangen-N\"urnberg,\\
Staudtstr.\, 7/B2, D-91058 Erlangen, Germany
}
\affiliation{Institute of Physics, University of Freiburg, Hermann-Herder-Strasse 3, D-79104 Freiburg, Germany}
\author{R.\ H\"artle}
\affiliation{Institut f\"ur Theoretische Physik, Friedrich-Hund-Platz 1, D-37077 G\"ottingen, Germany}
\affiliation{Institute of Physics, University of Freiburg, Hermann-Herder-Strasse 3, D-79104 Freiburg, Germany}
%
%
\author{M.\ Thoss}
\affiliation{Institut f\"ur Theoretische Physik und Interdisziplin\"ares Zentrum f\"ur Molekulare
Materialien, \\
Friedrich-Alexander-Universit\"at Erlangen-N\"urnberg,\\
Staudtstr.\, 7/B2, D-91058 Erlangen, Germany
}
\affiliation{Institute of Physics, University of Freiburg, Hermann-Herder-Strasse 3, D-79104 Freiburg, Germany}




\date{\today}

\begin{abstract}
We present a novel hierarchical quantum master equation (HQME) approach which provides a numerically exact description of nonequilibrium charge transport in nanosystems with electronic-vibrational coupling. In contrast to previous work [Phys.\ Rev.\ B {\bf 94}, 201407 (2016)], the active vibrational degrees of freedom are treated in the reservoir subspace and are integrated out. This facilitates applications to systems with very high excitation levels, for example due to current-induced heating, while properties of the vibrational degrees of freedom, such as the excitation level and other moments of the vibrational distribution function, are still accessible. The method is applied to a generic model of a nanosystem, which comprises a single electronic level that is coupled to fermionic leads and a vibrational degree of freedom. 
Converged results are obtained in a broad spectrum of parameters, ranging from the nonadiabatic to the adiabatic transport regime. 
We specifically investigate the phenomenon of vibrational instability, that is, the increase of current-induced vibrational excitation for decreasing electronic-vibrational coupling. The novel HQME approach allows us to analyze the influence of level broadening due to both molecule-lead coupling and thermal effects. Results obtained for the first two moments suggest that the vibrational excitation is always described by a geometric distribution in the weak electronic-vibrational coupling limit. 
  
%
\end{abstract}


\maketitle
\section{Introduction}
Electron transport in nanosystems, such as single-molecule junctions,\cite{Tal2008,Secker2011,Neel2011,Lau2016} nanoelectromechanical systems\cite{Craighead2000,Ekinci2005} as well as suspended carbon nanotubes,\cite{Weig2004,Sapmaz2006,Leturcq2009} has been the focus of many experimental and theoretical studies. Due to the small mass and size of these systems, the transport behavior is strongly influenced by the interplay between electronic and vibrational or structural degrees of freedom. This results in a variety of interesting transport phenomena such as multistability,\cite{Wilner2013} switching,\cite{Molen2010} negative differential resistance, \cite{Gaudioso2000,Pop2005,Galperin2005,Leijnse2008,Haertle2011} nonadiabatic effects,\cite{Reckermann2008,Frederiksen2008,Repp2010,Erpenbeck2015} enhanced current fluctuations\cite{Koch2005,Koch2006,Secker2011,Schinabeck2014} and decoherence,\cite{Haertle2011b,Ballmann2012} as well as local heating and cooling.\cite{Pop2005,Koch2006a,Haertle2011,Haertle2011c,Haertle2015a,Gelbwaser2017,Haertle2018} 
As the transport behavior is often governed by strong electronic-vibrational correlations\cite{Schinabeck2016} and pronounced nonequilibrium effects,\cite{Koch2005,Koch2006,Koch2006a,Haertle2011c,Schinabeck2014,Haertle2015a,Schinabeck2016,Gelbwaser2017,Haertle2018} the proper theoretical description is challenging. 

The effect of electronic-vibrational coupling on the transport properties of nanosystems has been studied by several approximate methods such as inelastic scattering theory,\cite{Bonca1995,Ness2001,Cizek2004,Cizek2005,Toroker2007,Benesch2008,Zimbovskaya2009} master equation approaches,\cite{May2002,Mitra2004,Lehmann2004,Koch2005,Koch2006,Haupt2006,Zazunov2006,Siddiqui2007,May2008,May2008a,Leijnse2008,Esposito2009a,Haertle2011,Volkovich2011,Schinabeck2014,Erpenbeck2016} nonequilibrium Green’s function (NEGF)\linebreak theory\cite{Flensberg2003,Mitra2004,Galperin2006,Ryndyk2006,Frederiksen2007,Haertle2008,Tahir2008,Haupt2009,Novotny2011,Haertle2011b,Utsumi2013,Erpenbeck2015} and functional renormalization group. \cite{Laakso2014,Khedri2017,Khedri2017a} Scattering theory methods typically neglect the nonequilibrium effects related to current-induced vibrational excitation.\cite{Haertle2008}
The other approaches also rely on approximations, which are usually perturbative in nature, employ decoupling or factorization schemes or other low order truncations.
On the one hand, these methods have provided profound insight into different transport phenomena. On the other hand, the predictive power of these approaches, in particular for strong coupling scenarios, is limited. Considering, e.g., the inelastic tunneling signal in the off-resonant transport regime for a vibrational mode in nonequilibrium, different approximate methods lead to contradictory predictions on the peak-dip transition of the first inelastic cotunneling feature. Basic NEGF methods predict the transition to occur at a zero bias conductance of half a conductance quantum.\cite{Galperin2004,Paulsson2005,Paulsson2006,Vega2006,Avriller2009,Schmidt2009,cuevasscheer2010}  
Nonequilibrium effects cause deviations from this prediction as shown by Novotny \emph{et al.}\cite{Novotny2011} and Utsumi \emph{et al.}.\cite{Utsumi2013} These deviations were confirmed by the hierarchical quantum master equation (HQME) formalism,\cite{Schinabeck2016} which represents a numerically exact method, i.e. a method which allows to systematically converge the results.

The HQME approach [also known as hierarchical equation of motion (HEOM) approach] is based on a partitioning into system and reservoir (in the following referred to as 'bath'). It generalizes perturbative master equation methods by including higher-order contributions as well as non-Markovian memory via a hierarchy of auxiliary density operators (ADOs). In particular, it allows for a systematic convergence of the results.\cite{Haertle2013a,Haertle2015} This approach was originally developed by Tanimura and Kubo\cite{Tanimura1989,Tanimura2006} to study relaxation dynamics in systems with a bosonic environment.\cite{Ishizaki2005,Schroeder2007,Chen2009a,Zhu2011,Kreisbeck2011,Struempfer2012,Liu2014,Tsuchimoto2015} Yan and co-workers\cite{Jin2008,Zheng2009,Li2012,Zheng2013,Cheng2015,Ye2016,Cheng2017,Li2017,Hou2017,Hou2017a} as well as H\"artle \emph{et al.}\cite{Haertle2013a,Haertle2014,Haertle2015,Wenderoth2016} have applied it to charge transport in quantum dot systems with electron-electron interactions.
Recently, an imaginary-time formulation\cite{Tanimura2014,Song2015,Tanimura2015} as well as a Wigner-space representation\cite{Sakurai2014,Tanimura2015} has been proposed. Other numerically exact methods to simulate vibrationally coupled charge transport in nanosystems include iterative path integral approaches,\cite{Huetzen2012,Simine2013,Weiss2013} diagrammatic quantum Monte Carlo simulations,\cite{Muehlbacher2008,Schiro2009,Han2010,Albrecht2012,Albrecht2015,Klatt2015} the numerical renormalization group technique,\cite{Jovchev2013,Jovchev2015,Khedri2017,Khedri2017a} the multilayer multiconfiguration time-dependent Hartree method. \cite{HWang2009,HWang2011,HWang2013,Wilner2013,Wilner2014,HWang2016}

In this paper, a hierarchical quantum master equation (HQME) approach is formulated to study vibrationally coupled charge transport. The method is particularly well suited to study scenarios with high transport-induced nonequilibrium vibrational excitation and, as such, is
complementary to our recently introduced HQME method for vibrationally coupled charge transport (see Ref.\ \onlinecite{Schinabeck2016}). 
Both approaches differ by the treatment of electronic-vibrational coupling: Within the HQME formalism introduced in this work, the vibrational degrees of freedom of the nanosystem are considered as part of the bath subspace (in the following referred to as VibBath). In contrast, the vibrations are treated as part of the reduced system (VibSys) in Ref.\  \onlinecite{Schinabeck2016}.
As a result, the method VibBath can more efficiently treat transport in systems, where a high nonequilibrium vibrational excitation occurs and thus a large vibrational basis set would be necessary within the approach VibSys. Although in the approach VibBath the vibrational degrees of freedom are treated  within the bath subspace and are integrated out, nonequilibrium effects are fully taken into account. This is in contrast to the approximate HQME method of Jiang \emph{et al.},\cite{Jiang2012} where, due to the polaron transformation employed, treating the vibrations and the leads in the bath subspace is equivalent to neglecting to a large extend the transport-induced nonequilibrium excitation of the vibration. Furthermore, properties such as the moments of vibrational distribution function can be accessed in the approach VibBath via the ADOs without any additional numerical effort. This is explicitly demonstrated for the first two moments.

We apply the approach VibBath to a generic model of vibrationally coupled charge transport. It comprises a single electronic level which is coupled to two macroscopic leads as well as a vibrational mode. 
We specifically consider the regime of small electronic-vibrational coupling $\lambda$.\cite{Mitra2004,Avriller2011,Haertle2011} In the limit $\lambda \to 0$, the average vibrational excitation can reach very high levels in the resonant transport regime and for sufficiently high bias voltages.\cite{Koch2006a,Haertle2011c,Kast2011,Haertle2015a} In molecular junctions, such high excitation levels could lead to the dissociation of the molecule.\cite{Gelbwaser2017} Therefore, this phenomenon is also referred to as vibrational instability.
Our work extends previous studies\cite{Koch2006a,Haertle2011c,Haertle2015a,Gelbwaser2017}, which were based on a lowest-order expansion in molecule-lead coupling and thus neglected the broadening of the electronic level by molecule-lead coupling.
H\"artle and Kulkarni\cite{Haertle2015a} showed that a finite lead temperature may result in a reduction of the average vibrational excitation compared to zero temperature and argued that the broadening due to molecule-lead coupling may have the same effect.
By means of numerically exact results, we check this conjecture and systematically study the influence of broadening induced by molecule-lead coupling on the vibrational distribution. 
These findings do not only apply to nanosystems with electronic-vibrational coupling but can be transferred to system with light-matter interactions such as a quantum dot which is coupled to a microwave cavity. \cite{Delbecq2011,Delbecq2013,Liu2014a,Cottet2015b} In the latter systems, the direct measurement of vibrational excitation is more feasible.\cite{Viennot2014} 

The paper is organized as follows: The model system and the HQME formalism are introduced in Secs.\ \ref{sec:model} and \ref{sec:theory_HEOM}.  
In Sec.\ \ref{sec:observables}, we outline how the observables of interest like the current and the nonequilibrium vibrational excitation can be obtained. The results are presented in Sec.\ \ref{sec:results}. First, it is demonstrated in Sec.\ \ref{sec:applicability} that the novel HQME approach can be applied in a broad range of parameters including (non-)adiabatic as well as (off-)resonant transport to obtain converged results for vibrationally coupled transport. This is accompanied by a discussion of the convergence properties. Second, in Sec.\ \ref{sec:vib_instab}, the transport-induced nonequilibrium vibrational excitation is investigated in the regime of small
electronic-vibrational coupling. Especially, the influence of molecule-lead coupling onto the vibrational excitation is studied on the basis of numerically exact HQME results. Sec.\ \ref{sec:conclusion} concludes. Throughout the paper, we use units where $\hbar=1$, $e=1$ and $k_\tB=1$. To be specific, we apply the terminology used in the context of charge transport in molecular junctions. However, other nanosystems with electronic-vibrational can also be described as mentioned above.

\section{Theoretical Methodology} \label{sec:theory}
\subsection{Model Hamiltonian} \label{sec:model}
In order to investigate vibrationally coupled electron transport in molecular junctions, we employ the following model Hamiltonian:\cite{Braig2003,Mitra2004,Cizek2004,Koch2005,Wegewijs2005,Haertle2008}
\begin{align}
 H = H_{\el} + H_\vib + H_{\el-\vib} + H_\leads + H_{\el-\leads}
\end{align}
with
\begin{subequations}
\begin{align}
H_{\el} = & \epsilon_0 d^{\dagger} d,\\
H_\vib = & \Omega a^{\dagger} a, \\
H_{\el-\vib} = & \lambda (a +a^\dagger) d^{\dagger} d, \label{eq:mol_vib_coup}\\
H_\leads =& \sum_{k \in \tL/\tR} \epsilon_{k} c_{k}^{\dagger} c_{k},\\
H_{\el-\leads} = &\sum_{k \in \tL/\tR} (V_{k} c_{k}^{\dagger} d +V_{k}^\ast d^{\dagger} c_{k}). \label{eq:mol_lead_coup}
\end{align}
\end{subequations}
A single electronic level with energy $\epsilon_0$ located on the molecular bridge is coupled to a continuum of electronic states in the macroscopic leads via interaction matrix elements $V_{k}$. The energy of these lead states is given by $\epsilon_{k}$. The operators $d^\dagger/d$ and $c_{k}^\dagger/c_{k}$ denote the creation / annihilation operators for the single electronic state on the molecular bridge and the states in the leads, respectively.
The interaction between the molecule and the left and the right lead, respectively, is characterized by the spectral densities (level width functions) $\Gamma_{\tL/\tR} (\omega)=2 \pi \sum_{k \in \tL/\tR} | V_k |^2 \delta (\omega-\epsilon_k)$.
The electrons transported through the molecular junction couple to the vibrational modes of the molecule. In this paper we consider a single vibrational mode described within the harmonic approximation with frequency $\Omega$ and corresponding creation and annihilation operators $a^\dagger/a$. The coupling strength between the vibrational mode and the electronic state is given by $\lambda$.

At this point, it is useful to employ a system-bath partitioning where the molecular energy level is considered as the reduced system, $H_\tS=H_{\el}$. The reduced system is coupled to two separate baths, the leads and the vibrational mode, $H_\tB=H_\leads+ H_\vib$, via the system-bath coupling $H_\tSB=H_{\el-\leads} + H_{\el-\vib}$. This partitioning also suggests to represent the Hamiltonian 
in the bath-interaction picture
\begin{align}
H^\tI (t) = H_\tS + \e^{\ii H_\tB t} H_\tSB \e^{-\ii H_\tB t} \equiv H_\tS + H^\tI_{\tSB} (t)
\end{align}
with
\begin{subequations}
\begin{align}
 H^\tI_{\tSB} (t) =& H^\tI_{\tSB,\leads} (t) + H^\tI_{\tSB,\vib} (t), \\
 H^\tI_{\tSB,\leads} (t) =& \sum_{K=\tL,\tR} \left( \tilde c^\dagger_K (t) d + d^\dagger \tilde c_K (t) \right), \\
 H^\tI_{\tSB,\vib} (t) =& d^\dagger d\; \tilde q(t)
\end{align}
\end{subequations}
and the lead index $K=\tL,\tR$. The bath coupling operators $\tilde c^{(\dagger)}_K (t)$ and $\tilde q(t) = \tilde a(t) + \tilde a^\dagger (t)$ are given by
\begin{subequations}
\begin{align}
\tilde c^\sigma_K (t) =& \e^{\ii H_\leads t} \left( \sum_{k \in K} V_{k} c^\sigma_{k} \right) \e^{-\ii H_\leads t} 
= \sum_{k \in K} V_{k} c^\sigma_{k} \e^{\sigma i \epsilon_k t},\\
\tilde a^s(t)=& \lambda \e^{\ii H_\vib t} a^s \e^{-\ii H_\vib t} = \lambda a^s \text{e}^{si\Omega t}
\end{align}
\end{subequations}
with $c_k^{-(+)} \equiv c_k^{(\dagger)}$, $a^{-(+)} \equiv a^{(\dagger)}$ and $\sigma,s=\pm$.
\subsection{HQME formalism for electronic-vibrational coupling} \label{sec:theory_HEOM}
The total system is described by the density operator $\rho_\tot$. The Liouville-von Neumann equation describes the time evolution of this operator 
\begin{equation}
\frac{\partial \rho_\tot(t)}{\partial t}=-\ii [H^\tI (t),\rho_\tot(t)]_-,
\label{eq:LvN}
\end{equation}
where $[A,B ]_-\equiv AB -BA$ denotes the commutator.
This equation is formally solved by
\begin{align}
 \rho_\tot(t)= U(t,0) \rho_\tot(0) U^\dagger (t,0),
\label{eq:red_dens_U} 
\end{align}
where $U(t,0)=\TT \expo{-\ii \int_0^t \dd \tau H^\tI (\tau)}$ denotes the time-ordered propagator in bath-interaction picture.

Following the original derivation of the HQME / HEOM approach, \cite{Tanimura1989,Tanimura1990,Tanimura2006,Jin2008,Tu2008} we employ the Feynman-Vernon influence 
functional formalism. It connects the reduced density matrix at time $t$ to the initial state of the total system at time $t=0$. \cite{Feynman1963,Weiss2008} To this end, 
it is assumed that the initial state factorizes, i.e.\ $\rho_\tot (t=0)=\rho (0) \rho_\tB (0)$ with $\rho_\tB (0)=\rho_\leads(0) \rho_\vib(0)$, where $\rho_\leads(0)$ and $\rho_\vib(0)$ denote the thermal equilibrium density operators of the non-interacting leads and the non-interacting vibration at temperatures $T_\leads$ and $T_\vib$
\begin{subequations}
\label{initialdensmat}
\begin{align}
 \rho_\leads(0) =& Z_\leads^{-1} \e^{-( H_\leads - \mu_\tL N_\tL - \mu_\tR N_\tR)/ T_\leads},\quad
 Z_\leads=\tr{\leads}{\e^{-( H_\leads - \mu_\tL N_\tL - \mu_\tR N_\tR)/ T_\leads}},
 \label{eq:rho_lead}\\
 \rho_\vib(0) =& Z_\vib^{-1} \e^{- H_\vib/ T_\vib}, \quad Z_\vib=\tr{\vib}{\e^{- H_\vib / T_\vib}}.  
 \label{eq:rho_vib}
\end{align}
\end{subequations}
Thereby, $N_{\tL / \tR}=\sum_{k \in \tL / \tR} c_k^\dagger c_k$ represent the occupation number operators of the left and the right lead. 
The chemical potentials $\mu_{\tL/\tR}$ are given by $\mu_\tL = \Phi/2$ as well as $\mu_\tR = -\Phi/2$, assuming a symmetric drop of 
the bias voltage $\Phi$ at the contacts.

It is noted that the chosen type of the initial state $\rho_\tB (0)$ (cf.\ Eq.\ (\ref{initialdensmat})) is crucial for the derivation of the HQME formalism used in this paper. 
Together with the non-interacting nature of the baths and the linear structure of the bath coupling 
operators $\tilde c^\sigma_K (t)$ and $\tilde q(t)$, it allows us to evaluate the Feynman-Vernon influence functional in closed form via 
Gaussian integration and, thus, to represent it in terms of the two-time correlation functions of the baths 
\begin{subequations}
\label{eq:bathcorfunc}
\begin{align}
C^\sigma_{K}(t-\tau) =& \tr{\leads}{\tilde c_K^\sigma(t) \tilde c_K^{\bar \sigma} (\tau) \rho_\leads(0)}, \label{eq:C_leads} \\
C_\vib(t-\tau) =& \tr{\vib}{\tilde q(t)\ \tilde q (\tau) \rho_\vib(0)}
\equiv C^-_\vib (t-\tau) + C^+_\vib (t-\tau) \label{eq:C_vib} \\
C^s_\vib(t-\tau) =& \tr{\vib}{\tilde a^s(t) \tilde a^{\bar s} (\tau) \rho_\vib(0)}
\label{eq:Cs_vib}
\end{align}
\end{subequations}
with $ \bar \sigma=-\sigma$ and $\bar s=-s$. The correlation functions in Eq.\ (\ref{eq:bathcorfunc}) are defined with respect to the initial state at $t=0$.

In the following, we outline the derivation of the Feynman-Vernon influence functional formalism, which is based on Eq.\ (\ref{eq:red_dens_U}).
A non-normalized coherent state is used as a basis for the reduced system $H_\tS=H_{\el}$
\begin{align}
 \ket{\Phi}=&\e^{- \Phi  d^\dagger} \ket{0}.
 \label{eq:f_coh_st}
\end{align}
This state is an eigenstate of the annihilation operator $d$ and thus fulfills the equation $d \ket{\Phi} =\Phi \ket{\Phi}$. The adjoint state $\bra{\Phi}$ is a left eigenstate of the corresponding creation operator $d^\dagger$ with eigenvalue $\Phi^*$. As the fermionic creation and annihilation operators obey anticommutation relations, the eigenvalues must also anticommute and are thus given by Grassmann variables. A detailed survey of the properties of fermionic coherent states and Grassmann variables can be found,  e.g., in Refs.\ \onlinecite{Negele2008,Ryder1996,Faddeev1991}.

Introducing the reduced density operator $\rho(t) =\tr{\tB}{\rho_\tot(t)}$ and tracing out the bath degrees of freedom in Eq.\ (\ref{eq:red_dens_U}) leads to
\begin{align}
\begin{split}
\rho(\Phi_f, \Phi'_{f},t) \equiv & \braket{\Phi_f |\rho(t)|\Phi'_{f}}\\
 =&\int \dd \Phi^*_i \dd \Phi_i \e^{-\Phi^*_i \Phi_i} \int \dd \Phi'^*_{i} \dd \Phi'_{i} \e^{-\Phi'^*_i \Phi'_i}\ J(\Phi_f, \Phi'_{f},t; \Phi_i, \Phi'_{i},0)\ \rho (\Phi_i, \Phi'_{i},0),
\end{split}
\end{align} 
where $J(\Phi_f, \Phi'_{f},t; \Phi_i, \Phi'_{i},0)$ denotes the coherent state representation of the Liouville-space propagator, which is given by the path integral expression
\begin{align}
J(\Phi_f, \Phi'_{f},t; \Phi_i, \Phi'_{i},0)=&\int_{ \Phi(0)=\Phi_{i}}^{\Phi^* (t)=\Phi^*_{f}} \DD [\Phi^* (t), \Phi(t)]
\int_{\Phi'(0)=\Phi'_{i}}^{\Phi'^* (t)=\Phi'^*_{f}} \DD [\Phi'^* (t), \Phi'(t)] \nonumber \\
& \times \expo{\ii S_\tS[\Phi,t]} \FF [\Phi, \Phi',t] \expo{-\ii S_\tS[\Phi',t]}.
\label{eq:Liou_prop}
\end{align}
The action $S_\tS[\Phi,t]$ is defined as
\begin{align}
S_\tS(\Phi;t)=& \Phi^* (t) \Phi (t) + \int_{0} ^{t} \dd \tau \left[\ii \Phi^* (\tau) \pd{\Phi(\tau)}{\tau} - H_\tS (\Phi(\tau)) \right],
\end{align}
where the first term originates from the fact that the coherent state is not normalized and the second term represents the classical action of the reduced system.

The Feynman-Vernon influence functional $\FF[\Phi, \Phi',t]$ contains all information on the system-bath coupling
and represents the central quantity of the HQME-formalism. It is defined as\cite{Makri1999}
\begin{align}
\begin{split}
 \FF [\Phi,\Phi',t]= \text{Tr}_\tB \left\{\TT \expo{-\ii \int_0^t \dd t' H^\tI_\tSB(\Phi^*(t'), \Phi (t'),t') }  \rho_\tB(0) \right.\\
 \left. \TT^{-1}\expo{+\ii \int_0^t \dd t' H^\tI_\tSB(\Phi'^*(t'), \Phi' (t'),t')   } \right\},
 \label{eq:IF}
 \end{split}
\end{align}
where in the interaction Hamiltonian $H^\tI_\tSB(t)$ the system operators $d$ / $d^\dagger$ have been replaced by the corresponding Grassmann fields 
\begin{align*}
H^\tI_\tSB(\Phi^*,\Phi,t)= \sum_{K} \left[ \tilde c^\dagger_K(t) \Phi(t) + \Phi^*(t) \tilde c_K (t) \right] + \Phi^*(t) \Phi(t) \tilde q(t).
\end{align*}
As the two separate baths, the leads and the vibration, interact only via the reduced system, the influence functional factorizes\cite{Zheng2012}
\begin{align}
   \FF[\Phi,\Phi',t]=\FF_{\leads}[\Phi,\Phi',t] \times \FF_{\vib}[\Phi,\Phi',t],
\label{eq:IF_factor}
\end{align}
where $\FF_{\leads}[\Phi,\Phi',t]$ corresponds to the coupling to the leads and $\FF_{\vib}[\Phi,\Phi',t]$ to the coupling to the vibration.
Since the bath coupling operators $\tilde c_K^{(\dagger)} (t)$ as well as $\tilde q (t)$ obey Gaussian statistics, both influence functionals can be obtained in closed form, e.g., by Gaussian integration in path integral picture\cite{Negele2008} or by a second order cumulant expansion\cite{Makri1999} of the exponents in Eq.\ (\ref{eq:IF}).
Consequently, the following expressions are obtained
\begin{subequations}
\begin{align}
\FF_\leads[\Phi, \Phi';t]=&\expo{-\ii \int_0^t \dd \tau \sum_{\sigma=\pm} \AA^{\bar \sigma} \left[\Phi(\tau),\Phi'(\tau) \right] \sum_{K,l} \BB_{K,\sigma,l} \left[ \tau,\Phi,\Phi' \right] },\\
\FF_\vib[\Phi, \Phi';t]=&\expo{-\ii \int_0^t \dd \tau \AA^\vib \left[\Phi(\tau),\Phi'(\tau) \right] \sum_{s=\pm} \BB^\vib_{s} \left[ \tau,\Phi,\Phi' \right] }.
\label{eq:IF_el}
\end{align}
\end{subequations}
with
\begin{subequations}
 \begin{align}
\AA^\sigma \left[ \Phi(t),\Phi'(t) \right] =&\Phi^\sigma (t)+ \Phi'^\sigma (t),\\
\sum_l \BB_{K,\sigma,l} \left[ t;\Phi,\Phi' \right]=&-\ii \left( \int_0^t \dd \tau C^\sigma_{K} (t-\tau) \Phi^\sigma (\tau) - \int_0^t \dd \tau C^{\bar \sigma, *}_{K} (t-\tau) \Phi'^\sigma (\tau) \right),\\
\AA^\vib [\Phi(t),\Phi'(t)]=&\Phi^*(t) \Phi(t) - \Phi'^*(t) \Phi'(t),\\
 \BB^\vib_{s} [t,\Phi,\Phi']=&-\ii \left( \int_0^t \dd \tau C^s_\vib (t-\tau) \Phi^* (\tau) \Phi (\tau) - \int_0^t \dd \tau C^{\bar s,*}_\vib (t-\tau) \Phi'^* (\tau) \Phi' (\tau) \right).
 \end{align}
 \label{eq:AB}
\end{subequations}
Hereby, $C^\sigma_{K}(t-\tau)$ and $C^s_\vib(t-\tau)$ denote the bath correlation functions of the noninteracting lead $K$ ($K=\tL,\tR$) and the vibration, respectively, as defined in Eq.\ (\ref{eq:bathcorfunc}).
The quantities $\AA^\sigma \left[ \Phi(t),\Phi'(t) \right]$ and $\BB_{K,\sigma,l} \left[ \tau,\Phi,\Phi' \right]$ with $\sigma = \pm$ are linear in the Grassman fields $\Phi^\sigma$ ($\Phi^{- (+)} \equiv \Phi^{(*)}$) and thus they also exhibit Grassmann properties. In contrast, $\AA^\vib [\Phi(t),\Phi'(t)]$ and $\BB^\vib_{s} [t,\Phi,\Phi']$ are bilinear in the Grassman fields and thus behave as ordinary complex numbers.
The index $l$ 
is associated with a representation of the two-time correlation functions $C_K^\sigma(t)$ and $C^s_\vib (t)$ (cf.\ Eqs.\ (\ref{eq:C_leads}) and (\ref{eq:Cs_vib})) by a sum over exponentials,\cite{Jin2008,Haertle2013a}
\begin{subequations}
\begin{align}
C^\sigma_{K}(t) \approx & \sum_{l=0}^{l_\text{max}} \eta_{K,l} \e^{-\gamma_{K,\sigma,l} t},  \label{eq:C_leads_exp}\\
C^s_\vib (t) =& \eta_s \e^{-\gamma_{s} t}. \label{eq:C_vib_exp}
\end{align}
\end{subequations}
This representation allows us to formulate a closed set of HQME because each term is self-similar with respect to time derivative, 
i.e.\ $\partial_t \e^{- \gamma t} = - \gamma \e^{- \gamma t}$. Note that an expansion of the bath correlation function 
in terms of Chebyshev polynomials\cite{Tian2012,Popescu2015,Popescu2016} or other complete sets of orthogonal functions\cite{Tang2015} 
is also possible but represents an alternative only for short timescales.

The amplitudes $\eta_{s}$ and frequencies $\gamma_{s}$ of the expansion (\ref{eq:C_vib_exp}) are easily obtained 
because $C^s_\vib (t)$ defined in Eq.\ (\ref{eq:Cs_vib}) is already in exponential form 
with
\begin{subequations}
\label{eq:gamma_eta_vib}
\begin{align}
 \gamma_\pm =& \mp \ii \Omega, \\
  \eta_+ =& \lambda^2 \bar n(0),\\
  \eta_- =& \lambda^2 (1 + \bar n(0)),
\end{align}
\end{subequations}
where $\bar n(0)= (\e^{\Omega/T_\vib} -1 )^{-1}$ denotes the initial thermal-equilibrium occupation of the vibration.

The amplitudes $\eta_{K,l}$ and frequencies $\gamma_{K,\sigma,l}$ of the expansion (\ref{eq:C_leads_exp}) 
are obtained from the spectral representation 
\begin{align}
  C^\sigma_{K} (t)=\frac{1}{2 \pi} \int_{-\infty}^\infty \dd \omega \e^{\sigma \ii \omega t} \Gamma_{K} (\omega) f [\sigma (\omega-\mu_K)],
\label{eq:C_FT}
\end{align}
which relates $C_K^\sigma (t)$ to the continuous spectral density in the leads $\Gamma_K (\omega)$ and the Fermi distribution $f(x)=\left( \expo{x/T_\leads} +1 \right)^{-1}$.
In the studies reported below, we employ the wide-band approximation by using a Lorentzian form, 
\begin{align}
  \Gamma_K (\omega)= \frac{\Gamma W^2}{(\omega-\mu_K)^2 +W^2},
 \label{eq:spec_dens}
\end{align}
with a very high value of the band width $W=10^4 \unit{eV}$. 
This implies hat the overall molecule-lead coupling strength is essentially 
independent of energy, i.e.\ $\Gamma_\tL = \Gamma_\tR=\Gamma$. Note that more complicated 
spectral densities can also be described within the approach, using, e.g., a Meier-Tannor parametrization.\cite{Meier1999} 
The Fermi distribution $f(x)$ is also approximated by a sum-over-poles scheme. In this work, the Pade decomposition is applied (cf. App. \ref{app:lead_correlation}), which exhibits a better convergence than, e.g., the Matsubara decomposition if no further truncation is employed.\cite{Haertle2013a} Recently, a combination of the Pade decomposition and a low-frequency logarithmic discretization scheme has been proposed as an extension for low temperatures.\cite{Ye2017} Alternatively, the decomposition in Eq.\ (\ref{eq:C_leads_exp}) could also be obtained by a direct fit of of the bath correlation function with exponentials.\cite{Duan2017}

As the explicit hierarchy construction has already been discussed in Refs.\ \onlinecite{Jin2007, Jin2008, Zheng2012} for the coupling to a fermionic bath and in Ref.\ \onlinecite{Xu2005,Xu2007,Jin2007,Zheng2012} for the coupling to a bosonic bath, here we only outline the differences and give the final set of EOMs.
The hierarchy construction starts by taking the derivative of the influence functional with respect to time,
\begin{align}
 \pd{}{t} \FF =& \left(\pd{}{t} \FF_\leads \right) \FF_\vib + \FF_\leads \left(\pd{}{t} \FF_\vib \right) = -\ii \sum_{K,\sigma,l} \AA^{\bar \sigma} \FF^{(1,0)}_{K,\sigma,l|}
 -\ii \AA^\vib \sum_{s=\pm} \FF^{(0,1)}_{|s},
\label{eq:dtF}
\end{align}
where the product rule has been applied due to the factorization of the influence functional. 
The quantities $\FF^{(1,0)}_{K,\sigma,l|}=\BB_{K,\sigma,l} \FF$ and $\FF^{(0,1)}_{|s} = \BB^\vib_s \FF$ denote auxiliary influence functionals.
The procedure continues by deriving EOMs for the auxiliary influence functionals $\FF^{(1,0)}_{K,\sigma,l|}$ and $\FF^{(0,1)}_{|s}$ and results finally in a hierarchy of coupled EOMs.
In general, purely electronic ($p \neq 0$, $q=0$) , purely vibrational ($p=0$, $q \neq 0$) and mixed ($q \neq 0$, $p \neq 0$) auxiliary influence functionals are defined by
\begin{align}
\FF^{(p,q)}_{j_p \cdots j_1|s_q \cdots s_1} = \BB_{j_p} \cdots \BB_{j_1} \BB^\vib_{s_q} \cdots \BB^\vib_{s_1} \FF,
\label{eq:AIF_mixed}
\end{align}
where the multi-index $j_\alpha=(K_\alpha,\sigma_\alpha,l_\alpha)$ $(\alpha=1, \ldots, p)$ is used for convenience. In the above expression, the order of the quantities $\BB_{j}$ is important because they are Grassmann variables as already outlined before. In contrast, the variables $\BB^\vib_{s}$ are ordinary complex numbers and so their order is arbitrary. For the sake of clarity, the two index sets are separated by a vertical bar in the subscript of the auxiliary influence functionals.
The mixed auxiliary influence functional $\FF^{(p,q)}_{j_p \cdots j_1|s_q \cdots s_1}$ is of ($2p$)th order in the molecule-lead coupling $V_k$ as well as of ($2q$)th order in the electronic-vibrational coupling $\lambda$.

In order to return to the operator level, auxiliary Liouville propagators $J^{(p,q)}_{j_p \cdots j_1|s_q \cdots s_1}$ are defined in analogy to Eq.\ (\ref{eq:Liou_prop}), where $\FF$ is replaced by $\FF^{(p,q)}_{j_p \cdots j_1|s_q \cdots s_1}$. With these auxiliary propagators, auxiliary density operators $\rho_{j_p \cdots j_1|s_q \cdots s_1}^{(p,q)}$ (ADOs) can be introduced via
\begin{align}
 \rho_{j_p \cdots j_1|s_q \cdots s_1}^{(p,q)} (t) = J^{(p,q)}_{j_p \cdots j_1|s_q \cdots s_1}\ \rho(0).
\label{eq:ADO_def}
\end{align}
Consequently, the following set of coupled EOMs is obtained
\begin{align}
\begin{split}
\dot \rho_{j_p \cdots j_1|s_q \cdots s_1}^{(p,q)} =&- \left(\ii \LL_\tS + \sum_{m=1}^p\gamma_{j_m} + \sum_{k=1}^q  \gamma_{s_k} \right) \rho_{j_p \cdots j_1|s_q \cdots s_1}^{(p,q)}\\
&-\ii \sum_{m=1}^p (-1)^{p-m} \CC_{j_m} \rho^{(p-1,q)}_{j_p \cdots j_{m+1} j_{m-1} \cdots j_1|s_q \cdots s_1} -\ii \sum_{k=1}^q \CC^\vib_{s_k} \rho^{(p,q-1)}_{j_p \cdots j_1|s_q \cdots s_{k+1} s_{k-1} \cdots s_1}\\
&-\ii \sum_{j=(K,\sigma,l)} \AA^{\bar \sigma} \rho^{(p+1,q)}_{jj_p \cdots j_1|s_q \cdots s_1} -\ii \AA^\vib \sum_{s=\pm} \rho^{(p,q+1)}_{j_p \cdots j_1|s s_q \cdots s_1},
\end{split}
\label{eq:EOM_vib}
\end{align}
where $\LL_\tS O = [H_\tS,O]_-$, $\rho \equiv \rho^{(0,0)}$ and $\rho^{(p<0,q)}=\rho^{(p,q<0)}=0$ hold.
The superoperators $\AA$, $\CC_{K,\sigma,l}$, $\AA^\vib$ and $\CC^\vib_{s}$ act in the following way
\begin{subequations}
\begin{align}
 \AA^{\bar \sigma} \rho^{(p,q)}=&d^{\bar \sigma} \rho^{(p,q)} + (-)^p \rho^{(p,q)} d^{\bar \sigma},\\
 \CC_{K,\sigma,l} \rho^{(p,q)} =& \eta_{K,l} d^\sigma \rho^{(p,q)} - (-)^p \eta^{*}_{K,l}\ \rho^{(p,q)} d^\sigma, \\
 \AA^\vib \rho^{(p,q)}=& d^\dagger d \rho^{(p,q)} - \rho^{(p,q)} d^\dagger d, \\
 \CC^\vib_s \rho^{(p,q)}=& \eta_s d^\dagger d \rho^{(p,q)} - \eta_{\bar s}^* \rho^{(p,q)} d^\dagger d,
\end{align}
\end{subequations}
with $d^+ \equiv d^\dagger$ and $d^- \equiv d$.

In this work, the coupled set of EOMs is directly solved for the steady state by setting $\dot \rho^{(p \geq 0,q \geq 0 )}=0$. By exploiting the hermiticity relation of the ADOs
 \begin{align}
 \rho^{(p,q),\dagger} _{\bar j_p \cdots \bar j_1|\bar s_q \cdots \bar s_1}=\rho^{(p,q)} _{j_1 \cdots j_p|s_q \cdots s_1} = (-1)^{\text{Int}(p/2)} \rho^{(p,q)} _{j_p \cdots j_1|s_q \cdots s_1},
\label{eq:herm_rel_elvib}
\end{align}
with $\bar j_\alpha = (K_\alpha, \bar \sigma_\alpha, l_\alpha)$,
the number of ADOs which has to be stored in memory can be reduced significantly and simultaneously 
linear dependencies are removed from the linear system of equations.
%
The derivation of this relation is given in App. \ref{sec:herm_rel_vibbath}.

For numeric evaluation, the finite but large electronic as well as the infinite vibrational hierarchy\cite{Zheng2012} have to be truncated. To this end, different truncation schemes have been proposed to terminate the hierarchy at the $n$th tier. These include a time-nonlocal truncation (chronological time ordering prescription) of the hierarchy, which amounts to setting all ADOs of the $(n+1)$th tier to zero.\cite{Yan2004} In contrast, within the partial time ordering prescription (referred to as time-local truncation), a Markovian approximation for the ADOs of the $n$th-tier is applied, so that they can be directly expressed by $(n-1)$th tier ADOs.\cite{Xu2005} If the reduced system dynamics within the Markovian approximation is additionally neglected, the ``terminator'' of Tanimura and coworkers is obtained.\cite{Tanimura1991}

In this work, we apply the time-local truncation to terminate the electronic hierarchy (cf.\ App.\ \ref{app:closing}). 
Employing a Markovian approximation of the $p$th-tier of the electronic hierarchy, the $p$th tier ADOs can be expressed by $(p-1)$th tier ADOs and thus they do not have to be considered as dynamic variables anymore. This offers the advantage that the numerical effort (memory consumption, CPU-time) is reduced to the level of a time-nonlocal truncation of the electronic hierarchy at the $(p-1)$-tier.
A priori it is not clear if the outcome of a calculation with time-local or time-nonlocal truncation at the $p$th-tier is closer to the converged result.\cite{Schroeder2007}
However, according to our experience, a time local truncation at the $p$th-tier of the electronic hierarchy always outperforms a time-nonlocal truncation at the $(p-1)$th-tier in the steady state regime, where the numerical effort is comparable.
Regarding the vibrational hierarchy, we apply the time-nonlocal truncation scheme. Due to the $\delta$-shaped vibrational spectral density, the vibrational correlations are only indirectly damped by the coupling to the leads and are thus very long-lived. Consequently, the situation is maximally non-Markovian. Therefore, a time-local truncation of the vibrational hierarchy employing a Markovian approximation is not the method of choice.

We want to stress that the vibrational distribution is only at the initial time, $t=0$, given by a thermal distribution corresponding to temperature $T_\vib$. 
At times $t \neq 0$, the complete transport induced vibrational nonequilibrium distribution is taken into account, which has evolved from the thermal equilibrium distribution at time $t=0$. This nonequilibrium distribution may differ from the initial thermal equilibrium distribution in all its moments, especially in the steady state limit. It is demonstrated in the next section that these moments of the vibrational nonequilibrium distribution can be extracted from the ADOs, which encode the deviations from the initial thermal distribution.

\subsection{Observables of interest} \label{sec:observables}
Observables of the system, such as the population or coherences of the electronic level, can be obtained in the usual way from the reduced density matrix. In contrast to many other methods used for open quantum systems, however, the HQME-formalism also allows direct access to properties of the bath via the ADOs.\cite{Jin2008,Zhu2012,Song2017}
Jin \emph{et al.} showed that the average transient current $\av{I_K (t)}=-\frac{\dd}{\dd t} \av{N_K (t)}$, which is given by the change of the average occupation number $\av{N_K(t)}$ in lead $K$, can be extracted from the purely electronic ADOs of the first tier\cite{Jin2008}
  \begin{align}
      \av{I_K (t)}=& \ii \avg{[N_K(t), H^\tI_\tSB (t)]_-}{\tS + \tB}=\ii \sum_l \tr{\tS}{ \left(d \rho_{K,+,l|}^{(1,0)}(t) - d^\dagger \rho_{K,+,l|}^{(1,0),\dagger}(t) \right) }.
  \end{align}
Furthermore, Shi and coworkers derived expressions for the expectation values of the powers of the collective bath coordinate (for our model system $\lambda (a+a^\dagger)$) and the related higher-order moments of the heat current in a nonequilibrium spin-boson system via a path integral derivation.\cite{Zhu2012,Song2017}

In the following, we demonstrate, how the average excitation of the vibrational mode $\av{n(t)} = \tr{\tS+ \tB}{n(t) \rho_\tot(t)}$ with $n(t)=a^\dagger (t) a(t)$ as well as the corresponding variance $\av{n^2(t)}-\av{n(t)}^2$ can be expressed by ADOs. As these quantities are properties of the vibrational bath only, they are given by purely vibrational ADOs. In order to find these relations, we follow an approach similar to that used by Jin \emph{et al.} for the current in Ref.\ \onlinecite{Jin2008}. The Liouville-von Neumann equation for the reduced density matrix is compared with the HQME to identify expressions for the first-tier ADOs of the vibrational hierarchy. Taking the trace over the bath degrees of freedom in Eq.\ (\ref{eq:LvN}) and substituting $H^\tI_{\tSB,\vib} (t)= d^\dagger d \sum_{s=\pm} \tilde a^s (t)$ with $\tilde a^s(t) = \lambda a^s(t) = \lambda a^s \e^{s \ii \Omega t}$, an EOM for the reduced density matrix is obtained
\begin{align}
\dot \rho (t)=&-\ii \LL_\tS \rho(t) - \ii \tr{\tB}{ [ H^\tI_{\tSB, \leads} (t) ,\rho_\tot (t)]_- }
-\ii \left[ d^\dagger d, \sum_{s} \tr{\tB}{ \tilde a^s (t) \rho_\tot (t) } \right]_-.
\end{align}
By comparing this equation with the EOM for the reduced density matrix in the HQME-framework
\begin{align}
 \dot \rho (t)=& -\ii \LL_\tS \rho (t) -\ii \sum_j \AA^{\bar \sigma} \rho^{(1,0)}_{j|} -\ii \left[ d^\dagger d, \sum_{s} \rho^{(0,1)}_{|s} (t) \right]_-,
\end{align}
the relation
\begin{align}
 \rho^{(0,1)}_{|+} (t) + \rho^{(0,1)}_{|-} (t)=\tr{\tB}{\tilde a^+ (t) \rho_\tot (t)} + \tr{\tB}{\tilde a^- (t) \rho_\tot (t)}.
\end{align}
can be established, which suggests the identity
\begin{align}
 \rho^{(0,1)}_{|s} (t)=\tr{\tB}{\tilde a^s (t) \rho_\tot (t)}= \lambda\ \tr{\tB}{a^s (t) \rho_\tot (t)}.
 \label{eq:app_1bADO}
\end{align}
Comparing this expression with the definition of the average vibrational excitation \linebreak $\av{n(t)} = \tr{\tS+ \tB}{a^\dagger (t) a(t) \rho_\tot(t)}$, we can deduce that the vibrational excitation has to be related to purely bosonic ADOs of the second tier ($p=0, q=2$).
In order to find this relation and to confirm the conjecture in Eq.\ (\ref{eq:app_1bADO}), an EOM for the ADOs of the first vibrational tier is formulated by taking the time derivative of $\tr{\tB}{\tilde a^{s_1} (t) \rho_\tot (t)}$:
\begin{align}
 \begin{split}
 \frac{\dd}{\dd t} \tr{\tB}{\tilde a^{s_1} (t) \rho_\tot (t)} = & \tr{\tB}{ \dot{\tilde a}^{s_1} (t) \rho_\tot (t)} + \tr{\tB}{\tilde a^{s_1} (t) \dot \rho_\tot (t)} \\
= &-\ii \left( \LL_\tS - s_1 \Omega \right)  \rho^{(0,1)}_{s_1} (t) - \ii \tr{\tB}{ \tilde a^{s_1} (t) \left[ H^\tI_{\tSB, \leads} (t) ,\rho_\tot (t) \right]_- }\\
 & -\ii \sum_{s} \left( d^\dagger d\ \tr{\tB}{ \tilde a^{s_1} (t) \tilde a^s (t) \rho_\tot (t)}
 - \tr{\tB}{ \tilde a^s (t) \tilde a^{s_1} (t) \rho_\tot (t)} d^\dagger d \right)
\end{split}
\label{eq:as1}
\end{align}
with $\tr{\tB}{ \tilde a^{s_1} (t) \tilde a^s (t) \rho_\tot (t)}=\lambda^2 \e^{\ii (s_1 + s) \Omega t} \tr{\tB}{ a^{s_1} a^s  \rho_\tot (t)}$.
According to Eq.\ (\ref{eq:app_1bADO}), this result has to be compared to the EOM for the first purely vibrational ADO ($p=0, q=1$) in the HQME framework, which reads
\begin{align}
 \begin{split}
\dot \rho^{(0,1)}_{|s_1}=& - (\ii \LL_\tS + \gamma_{s_1}) \rho^{(0,1)}_{|s_1} - \ii \left( \eta_{s_1} d^\dagger d \rho - \eta_{\bar s_1} \rho d^\dagger d \right)\\
 & -\ii \sum_{K,l,\sigma} \left( d^{\bar \sigma} \rho^{(1,1)}_{K,\sigma,l|s_1} - \rho^{(1,1)}_{K,\sigma,l|s_1} d^{\bar \sigma} \right)
 - \ii \sum_s \left( d^\dagger d \rho^{(0,2)}_{|s s_1} - \rho^{(0,2)}_{|s s_1} d^\dagger d \right).
\end{split}
\label{eq:rho1}
\end{align}
As $\gamma_{s_1} = -s_1 \ii \Omega$ holds according to Eq.\ (\ref{eq:gamma_eta_vib}), the coefficients in front of $\rho^{(0,1)}_{|s_1}$ agree in Eqs.\ (\ref{eq:as1}) and (\ref{eq:rho1}), which confirms the assumption in Eq.\ (\ref{eq:app_1bADO}).
Additionally, the following terms have to be equal
\begin{subequations}
\begin{align}
 d^\dagger d \sum_{s} \tr{\tB}{ \tilde a^{s_1} (t) \tilde a^s (t) \rho_\tot (t) } \equiv\ & d^\dagger d \left( \eta_{s_1} \rho(t) + \sum_{s} \rho^{(0,2)}_{|s s_1} (t) \right) \label{eq:app_vibex1a} \\
\sum_{s} \tr{\tB}{ \tilde a^{s} (t) \tilde a^{s_1} (t) \rho_\tot (t) } d^\dagger d \equiv\ & \left( \eta_{\bar s_1} \rho (t) + \sum_{s} \rho^{(0,2)}_{|s s_1}(t) \right) d^\dagger d \label{eq:app_vibex1b}
\end{align}
\end{subequations}
Evaluating Eq.\ (\ref{eq:app_vibex1a})  for $s_1=+$ and Eq.\ (\ref{eq:app_vibex1b}) for $s_1=-$ leads to
\begin{subequations}
\begin{align}
 \lambda^2 \left( \tr{\tB}{ a^\dagger a \rho_\tot (t) } + \tr{\tB}{ a^\dagger a^\dagger \rho_\tot (t) } \right) =& \eta_+ \rho (t) +  \rho^{(0,2)}_{|+ +} (t) +  \rho^{(0,2)}_{|- +} (t) \label{eq:app_vibex2a}\\
 \lambda^2 \left( \tr{\tB}{ a^\dagger a \rho_\tot (t) } + \tr{\tB}{ a a \rho_\tot (t) } \right) =& \eta_+ \rho(t) +  \rho^{(0,2)}_{|- -} (t) +  \rho^{(0,2)}_{|- +} (t) \label{eq:app_vibex2b}
\end{align}
\end{subequations}
where we have used the fact that the order of the indices of bosonic ADOs is arbitrary.
Comparing Eqs.\ (\ref{eq:app_vibex2a}) and (\ref{eq:app_vibex2b}), the following relation is found
\begin{align}
 \tr{\tB}{ a^\dagger a \rho_\tot (t)} \equiv \frac{1}{\lambda^2} \left( \eta_+ \rho(t) +  \rho^{(0,2)}_{|- +} (t)  \right).
\end{align}
As a result, the average vibrational excitation is given by
\begin{align}
  \av{n(t)} = \tr{\tS + \tB}{a^\dagger a \rho_\tot (t)} =& \frac{1}{\lambda^2} \left( \eta_+ +  \rho^{(0,2)}_{|- +;00} (t) + \rho^{(0,2)}_{|- +;11} (t)  \right) \nonumber\\
  =& \bar n(0) + \frac{1}{\lambda^2} \left( \rho^{(0,2)}_{|- +;00}(t) + \rho^{(0,2)}_{|- +;11}(t)  \right), \label{eq:av_Vibex}
\end{align}
where $\bar n(0)$ denotes the initial average vibrational excitation of the vibrational mode at time $t=0$, given by the thermal distribution in Eq.\ (\ref{eq:rho_vib}).

In order to evaluate the variance of the vibrational excitation $\av{n^2(t)}-\av{n(t)}^2$, the second moment $\av{n^2(t)}$ has to be expressed in terms of ADOs. This can be easily achieved by extending the procedure used to determine $\av{n(t)}$ up to $\rho^{(0,4)}_{|s_4 s_3 s_2 s_1}$. As a result, the following relation is found
\begin{align}
\begin{split}
 \tr{\tS+\tB}{ a^\dagger a a^\dagger a \rho_\tot (t) } =& (2 \bar n(0)^2 + \bar n(0)) + \frac{1}{\lambda^2}( 4 \bar n(0) + 1) \left( \rho^{(0,2)}_{|- +;00}(t) + \rho^{(0,2)}_{|- +;11}(t)  \right)\\
 &+ \frac{1}{\lambda^4} \left( \rho^{(0,4)}_{|--++;00} (t) + \rho^{(0,4)}_{|--++;11} (t)\right),
\end{split}
\end{align}
where the term $(2 \bar n(0)^2 + \bar n(0))$ corresponds to the second moment of the initial thermal distribution. The other terms represent corrections due to the nonequilibrium vibrational excitation.

This relation has several implications: (i) In order to include the nonequilibrium excitation of the vibrational mode, at least the second vibrational tier of the hierarchy has to be included. (ii) Furthermore, the nonequilibrium vibrational distribution arises by corrections with respect to the initial thermal equilibrium distribution of the vibrational mode, i.e. the closer the initial distribution is to the final one, the smaller the corrections are which have to be provided by the hierarchy of equations.
The latter finding can be utilized to improve the convergence of the method in the steady state regime. As the steady state solution is assumed to be unique, the choice of the initial state is arbitrary in general. However, within the HQME formalism presented in this work, the initial state has to be a thermal state which is completely characterized by its average $\bar n(0)$ (geometric distribution).
In order to obtain stable and converged results with the minimum amount of tiers of the vibrational hierarchy, it is beneficial to set the initial average vibrational excitation $\bar n(0)$ as close as possible to its final nonequilibrium value. If we are interested in observables as a function of bias voltage, this requirement can easily be fulfilled: The average nonequilibrium vibrational excitation (cf.\ Eq.\ (\ref{eq:av_Vibex})) obtained as output for the past bias value is taken as input for the initial excitation for the actual bias value.
%
%

%
\section{Results} \label{sec:results}
In this section, the HQME approach VibBath introduced in Sec.\ \ref{sec:theory} is used to investigate vibrationally coupled electron transport in single-molecule junctions. First, we demonstrate in Sec.\ \ref{sec:applicability} the performance of the HQME method VibBath and show that it can be applied in a broad parameter space in order to obtain numerically exact results. The parameters range from the nonadiabatic to the adiabatic transport regime and from weak to strong electronic-vibrational coupling.
Second, the influence of the molecule-lead coupling $\Gamma$, which induces a broadening of the electronic level, on the vibrational excitation of the molecular bridge is investigated in Sec.\ \ref{sec:vib_instab}. We especially focus on the regime of weak electronic-vibrational coupling, which is governed by the counter-intuitive phenomenon that the vibrational excitation increases with decreasing electronic-vibrational coupling.%
%
%
\subsection{Performance of the HQME approach VibBath} \label{sec:applicability}
To demonstrate the performance of the HQME approach VibBath, we apply it to models of vibrationally coupled electron transport in different parameter regimes.
Specifically, the convergence properties with respect to the truncation of the electronic as well as the vibrational hierarchy are discussed. Additionally, the application range of the method VibBath is compared with the approach VibSys introduced in Ref.\ \onlinecite{Schinabeck2016}.

\begin{figure*}[h!]
\begin{tabular}{cc}
\includegraphics[width=0.45\textwidth]{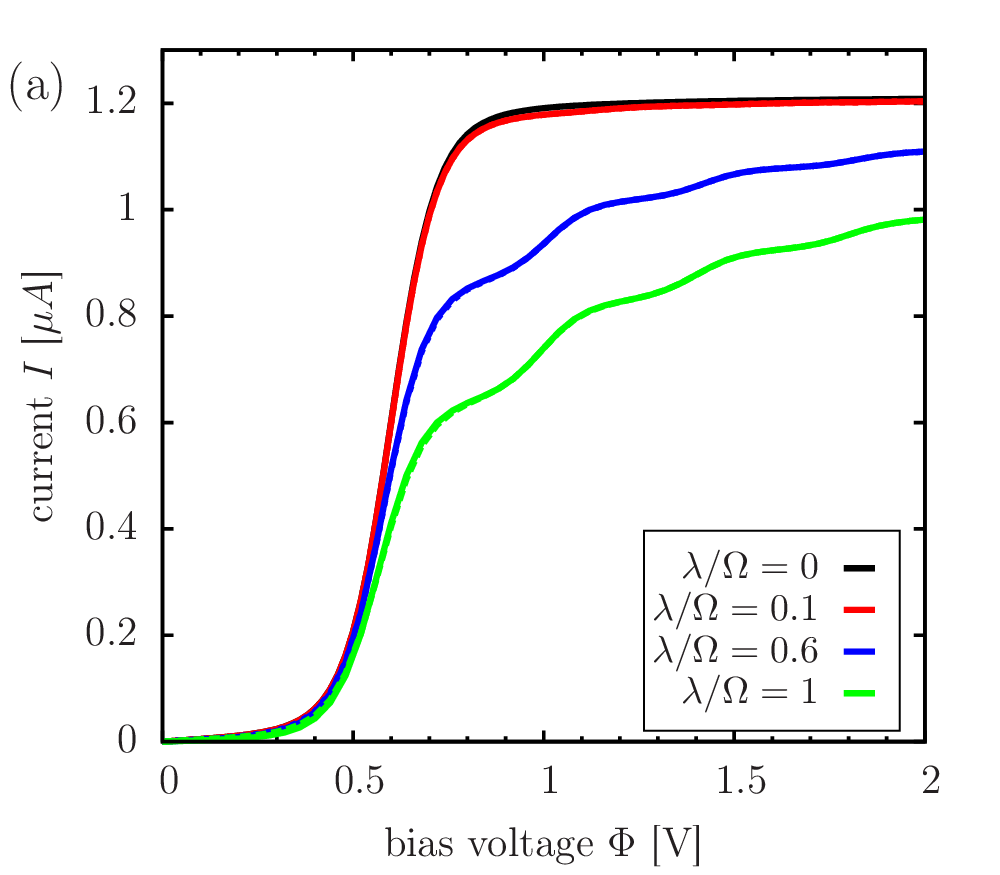}
\includegraphics[width=0.45\textwidth]{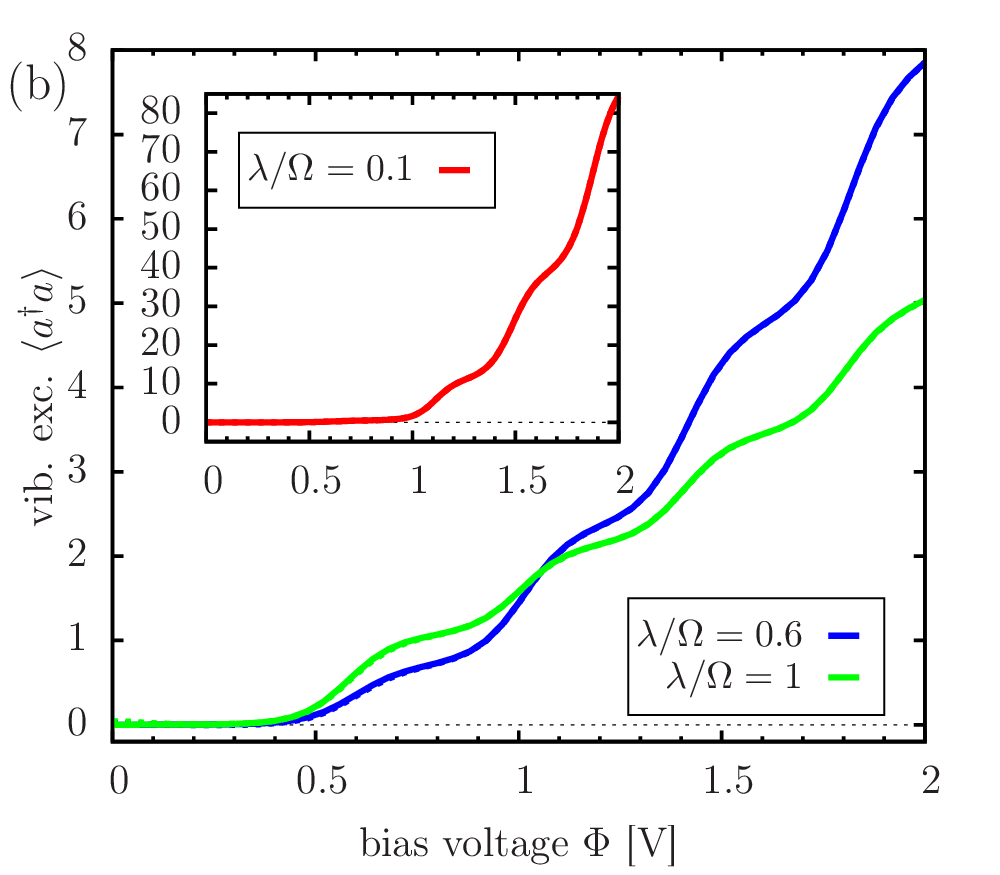}\\
%
\includegraphics[width=0.45\textwidth]{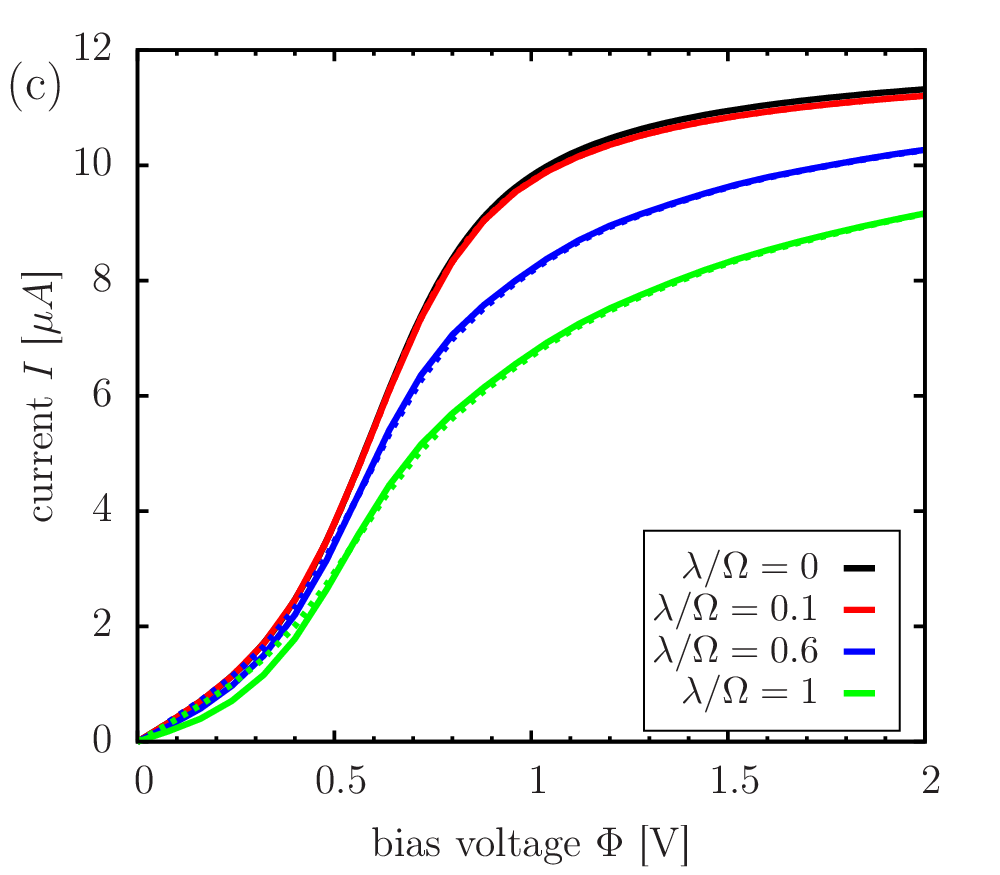}
\includegraphics[width=0.45\textwidth]{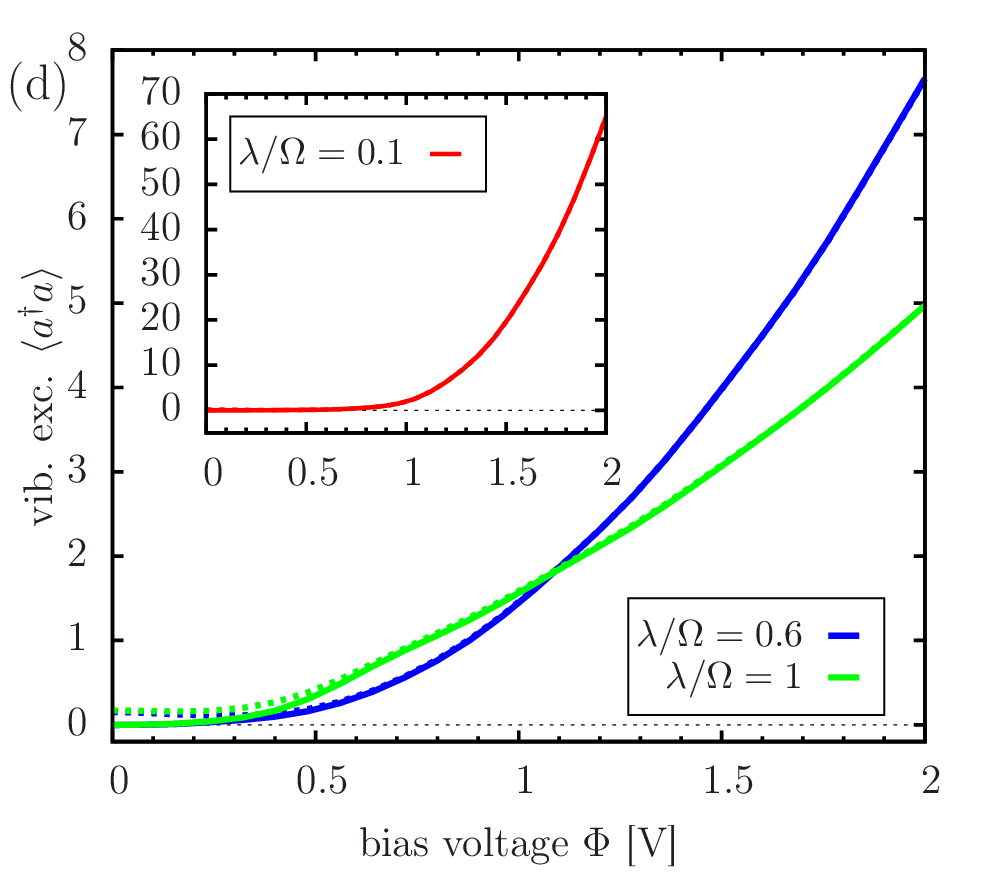}\\
%
\includegraphics[width=0.45\textwidth]{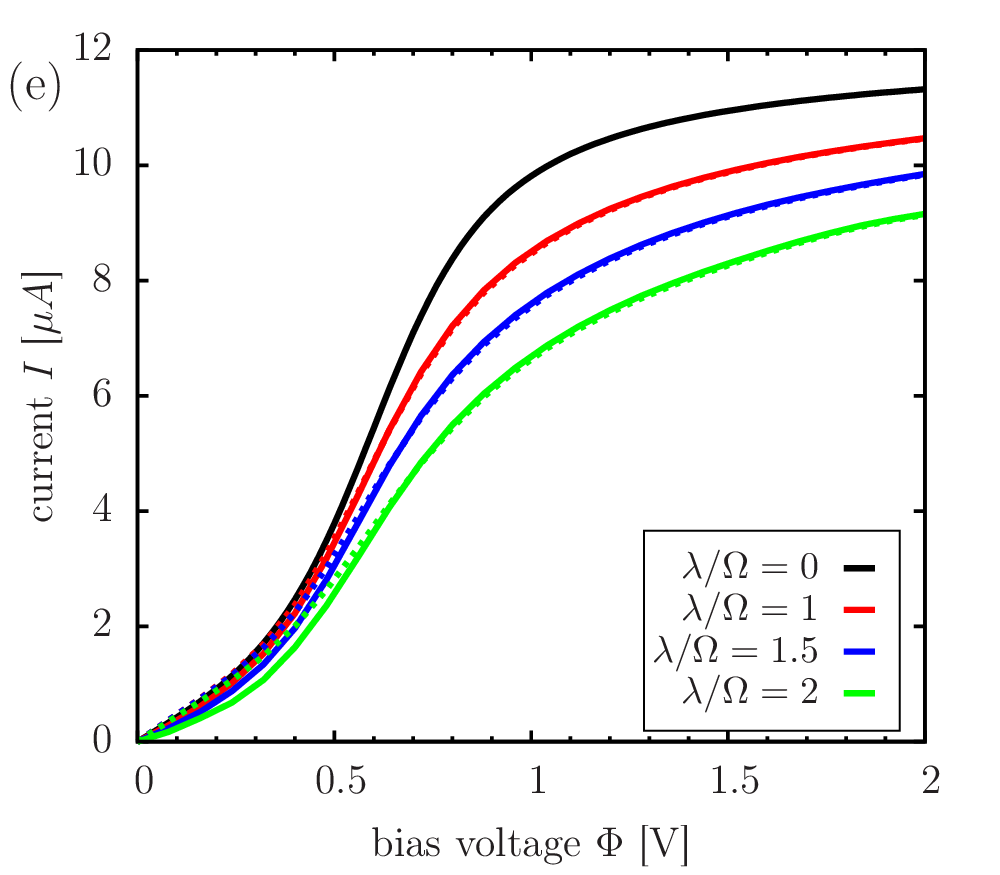}
\includegraphics[width=0.45\textwidth]{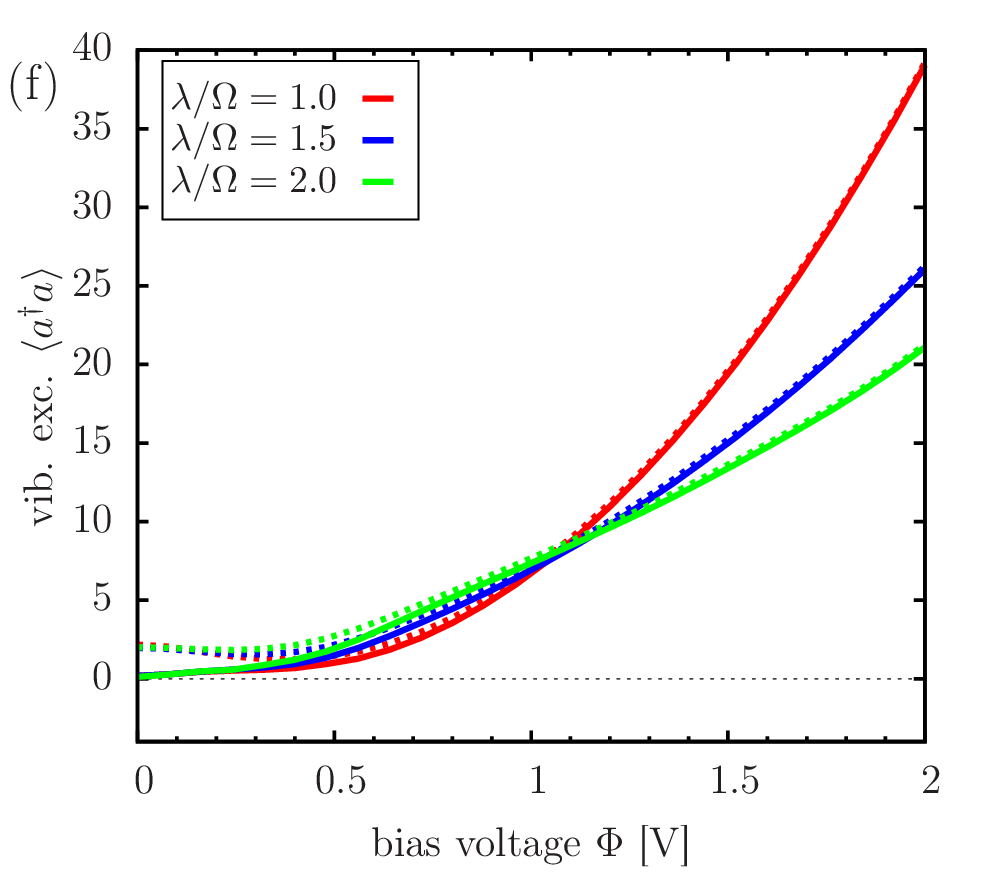}\\
\end{tabular}
\caption{Current-voltage (left) and vibrational excitation-voltage (right) characteristics for different electronic-vibrational couplings $\lambda/\Omega$. The results are obtained using the approach VibBath for model 1 (cf.\ Tab.\ \ref{tab:models}) and $\Gamma=0.01 \unit{eV}$ (a,b) as well as $\Gamma=0.1 \unit{eV}$ (c,d), and model 2 and $\Gamma=0.1 \unit{eV}$ (e,f). The solid (dashed) lines correspond to a time-local truncation of the electronic hierarchy after the third (second) tier.}
\label{fig:I_vib_V}
\end{figure*}

Fig.\ \ref{fig:I_vib_V} presents the current as well as the average vibrational excitation as a function of bias voltage. The parameters of the models considered are summarized in Tab.\ \ref{tab:models}.
\begin{table}[htb!]
 \centering
  \begin{tabular}{c | c c c c }
  Model & $ \tilde \epsilon_0$ [eV] & $\Omega$ [eV] & $T_\leads$ [K] & $\Phi$ [V] \\ \hline
  1 & 0.3 & 0.2 & 300 & variable \\
  2 & 0.3 & 0.05 & 300 & variable \\
  3 & 0.3 & 0.1 & variable & 0.9
 \end{tabular}
 \caption{Summary of model parameters where $\tilde \epsilon_0 = \epsilon_0 -\lambda^2/\Omega$ denotes the reorganized energy level.}
 \label{tab:models}
\end{table}
They range from the nonadiabatic transport regime in panels a and b (model 1 and $\Gamma=10^{-2} \unit{eV}$) to the adiabatic regime in panels e and f (model 2 and $\Gamma=0.1 \unit{eV}$) via the crossover regime in panels c and d (model 1 and $\Gamma=0.1 \unit{eV}$) for different electronic-vibrational couplings.
The solid lines are obtained by the HQME approach VibBath employing a truncation of the electronic hierarchy after the third tier with a time-local closure (cf.\ App. \ref{app:closing}). The respective truncation levels of the vibrational hierarchy are summarized in Tab.\ \ref{tab:vib_tiers}.
\begin{table}[htb!]
 \centering
  \begin{tabular}{c | c | c  | c | c }
  $\Omega$ [eV] & $ \Gamma$ [eV] & $ (\Gamma_{\rm L}+\Gamma_{\rm R}) / \Omega$ & $ \lambda / \Omega$  &  \# vibrational tiers \\ \hline
   & & & 0.1 & 10\\ \cline{4-5}
  0.2 & 0.01 & 0.1 & 0.6 & 24\\ \cline{4-5}
   & & & 1.0 & 30\\ \hline
  & & & 0.1 & 6\\ \cline{4-5}
  0.2 & 0.1 & 1 & 0.6 & 12\\ \cline{4-5}
  & & & 1.0 & 18\\ \hline
  & & & 1.0 & 24\\  \cline{4-5}
  0.05 & 0.1 & 4 & 1.5 & 27\\ \cline{4-5}
  & & & 2.0 & 30\\ \hline
 \end{tabular}
 \caption{Number of tiers of the vibrational hierarchy used in the calculations.}
 \label{tab:vib_tiers}
\end{table}
The results are converged with respect to the various numerical parameters as demonstrated in App.\ \ref{app:conv_prop}.

\begin{figure*}[h!]
	\centering
\begin{tabular}{llllllll}
(a)\hspace{0cm} & & (b)\hspace{0cm} & & (c)\hspace{0cm}  && (d) \hspace{0cm} &\\
& 
\includegraphics[width=.2\textwidth]{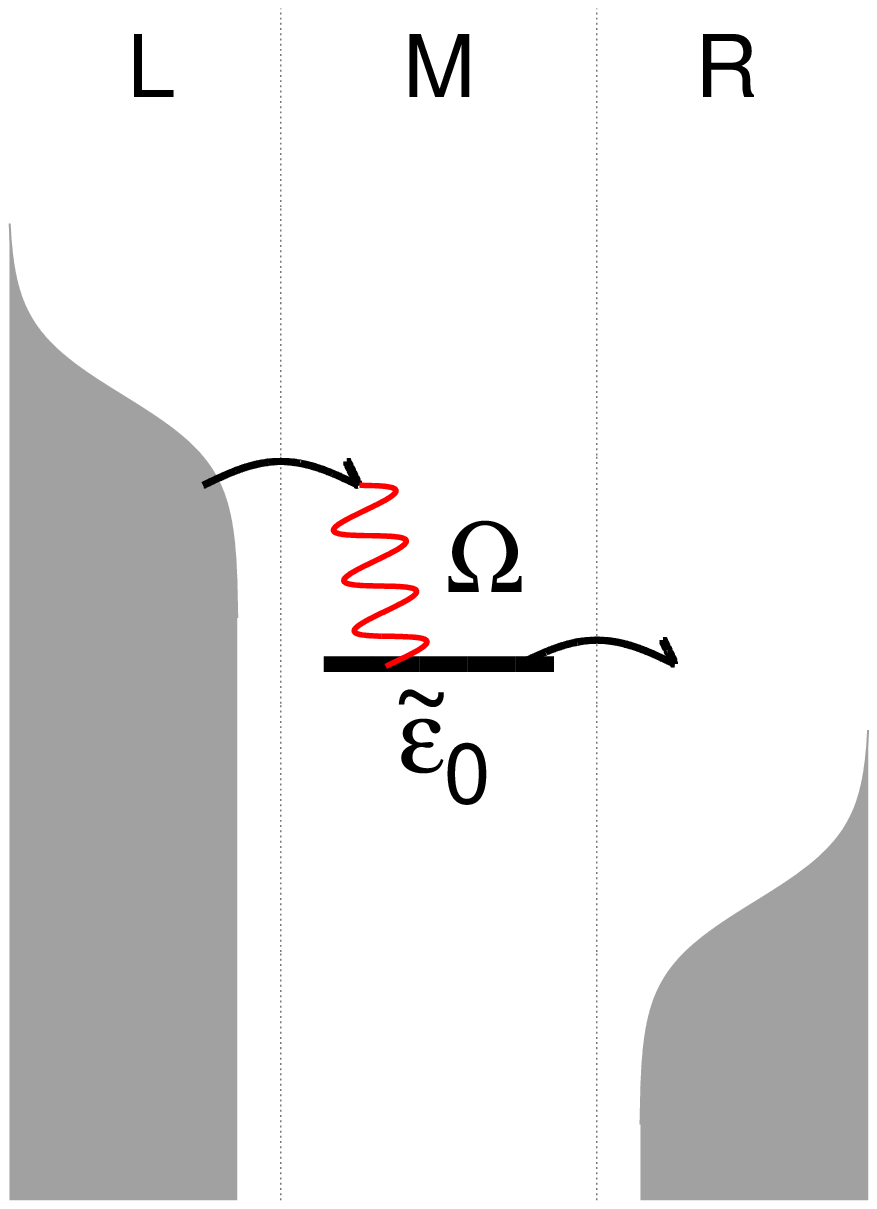}
& &
\includegraphics[width=.2\textwidth]{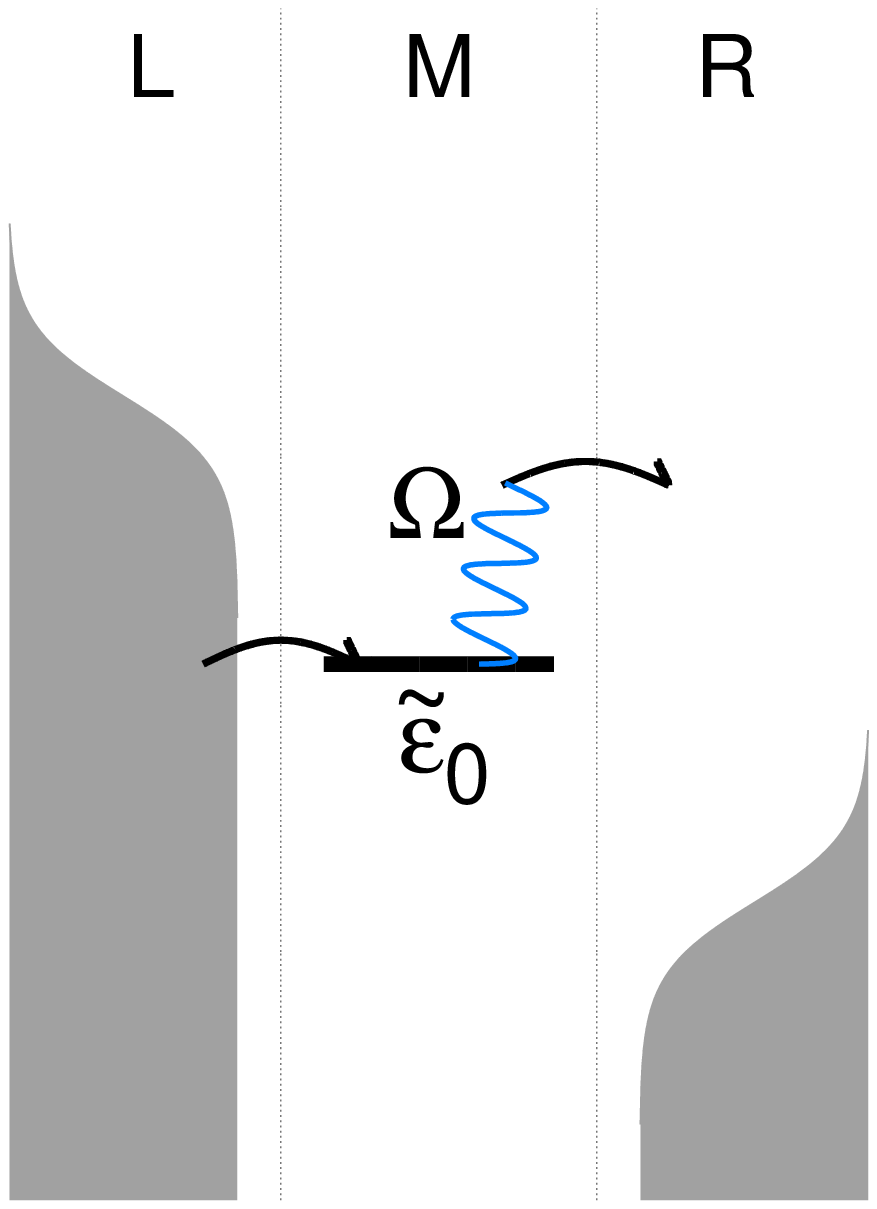}
& &
\includegraphics[width=.2\textwidth]{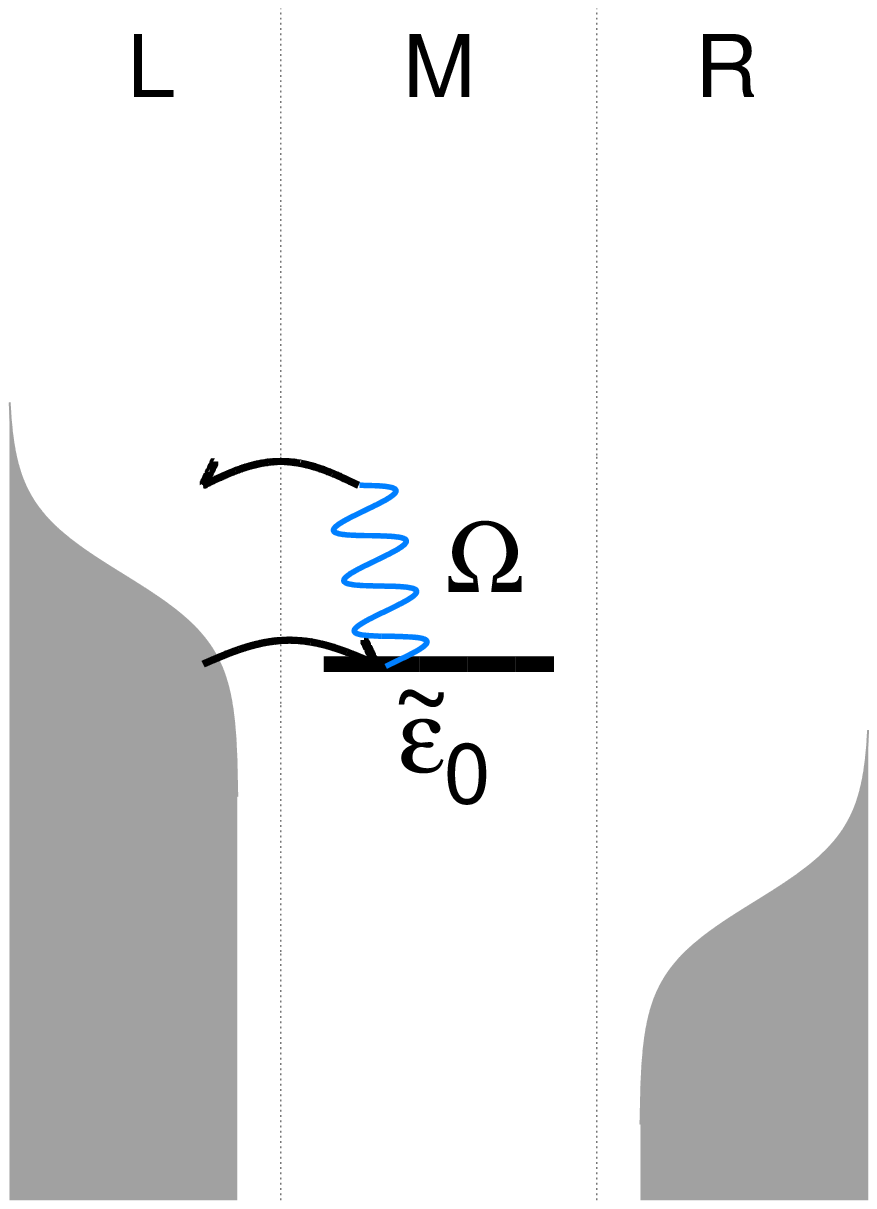}
& &
\includegraphics[width=.2\textwidth]{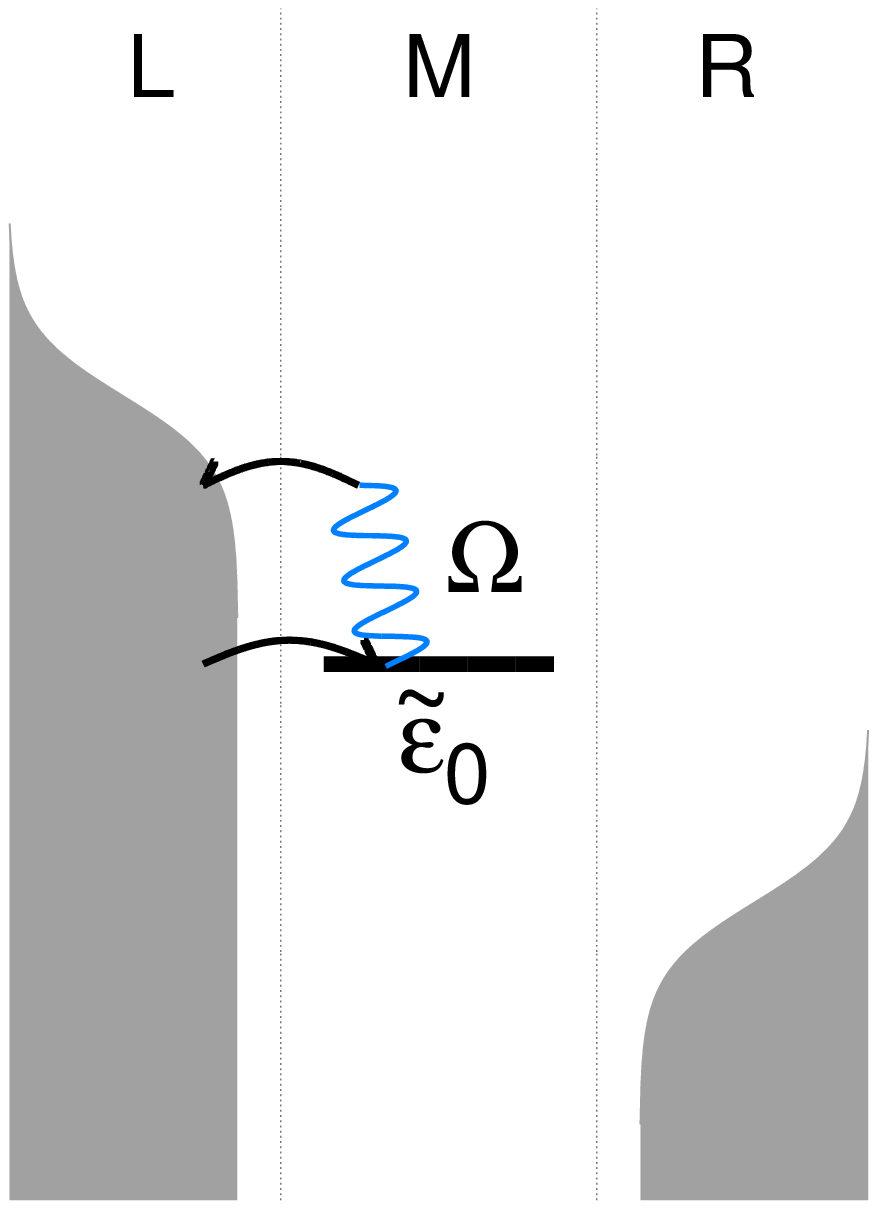}
\end{tabular}
		\caption{Examples of inelastic processes in the resonant transport regime at nonzero temperature of the leads. The shorthands L and R denote the left and the right lead, respectively, and M stands for the molecular bridge. Panel (a) (panel (b)) depicts an inelastic transport process where the vibrational mode is excited (deexcited) by a single vibrational quantum of frequency $\Omega$ [red (blue) wiggly line], while an electron sequentially tunnels from the left lead onto the molecular bridge and further to the right lead. Deexciting the vibration by a single quantum, an electron-hole pair can be created in the left lead, as demonstrated in panel (c). In panel (d), the same electron-hole pair creation process is suppressed by the increased chemical potential in the left lead ($\mu_\tL > \tilde \epsilon_0 + \Omega$), i.e.\ it is only enabled by the broadening of the Fermi distribution at finite temperature.}
\label{fig:sample_processes}
\end{figure*}

%
Figs.\ \ref{fig:I_vib_V}a and b show results for the current-voltage characteristics as well as the vibrational excitation obtained in the nonadiabatic transport regime ($2 \Gamma/\Omega=0.1$). The results in this regime
exhibit the typical Franck-Condon (FC) step structure, 
where the steps correspond to the opening of inelastic transport channels. Examples for such processes are depicted in Figs.\ \ref{fig:sample_processes}a,b.
For small electronic-vibrational coupling, $\lambda/\Omega=0.1$, diagonal FC-transitions are dominant, i.e. the transport is governed by elastic processes and thus the step structure in the current is barely visible.
With increasing coupling, the current is suppressed at the onset of the resonant transport regime ($\Phi \gtrsim 2 \tilde \epsilon_0$) because the FC transition probability between the vibrational ground state of the occupied and unoccupied molecular bridge is reduced.
The vibrational excitation exhibits a more complex dependence on the strength of the electronic-vibrational coupling. While for smaller voltages the excitation increases with coupling, at larger voltages this dependence reverses. The increase at smaller voltages is due to the fact that stronger coupling favors excitations comprising a larger number of vibrational quanta, thus leading to a higher vibrational excitation.
The reversed behavior at larger voltages can be attributed to the fact that the vibrational excitation is not only influenced by transport-related processes (cf.\ Fig.\ \ref{fig:sample_processes}a,b) but also by resonant electron-hole pair creation processes (cf.\ Fig.\ \ref{fig:sample_processes}c),\cite{Haertle2011,Haertle2011c,Haertle2015a} which for the lower temperatures ($T \ll \Omega $) considered in Fig.\ \ref{fig:I_vib_V} result  in a deexcitation of the vibrational mode. In contrast to the transport-induced processes,  deexcitation induced by electron-hole pair creation is blocked for larger voltages (cf.\ Fig.\ \ref{fig:sample_processes}d). Deexcitation processes involving transitions with a smaller number of vibrational quanta, which dominate for weak electronic-vibrational coupling, are blocked first. As a result of these missing cooling processes, vibrational excitation increases with decreasing coupling for larger voltages.

For larger molecule-lead coupling ($2 \Gamma=\Omega$), the step structures in the current- and vibrational excitation-voltage characteristics, which are depicted in Figs.\ \ref{fig:I_vib_V}c,d, is smoothed by the increased broadening of the electronic level due to molecule-lead coupling.

Increasing the molecule-lead coupling further, the adiabatic transport regime is entered. As an example, Figs.\ \ref{fig:I_vib_V}e,f show results for model 2 ($\Omega=0.05 \unit{eV}$) and $2\Gamma/\Omega=4$.
In this regime, the method VibBath allows to describe systems with strong electronic-vibrational coupling ($\lambda/\Omega >1$). Due to the reduced frequency (energy) of the vibration, the average vibrational excitation for $\lambda/\Omega=1$ in Fig.\ \ref{fig:I_vib_V}f is significantly higher than in Fig.\ \ref{fig:I_vib_V}d.

Next, we discuss the convergence of the newly developed approach VibBath with respect to the truncation of the electronic and the vibrational hierarchy on the basis of Fig.\ \ref{fig:I_vib_V}. 
The number of tiers of the electronic hierarchy, which has to be included to reach convergence, is typically determined by the molecule-lead coupling strength $\Gamma$ and the temperature $T$ of the leads.\cite{Li2012,Haertle2013a,Haertle2015} The higher $\Gamma$ is the more tiers of the electronic hierarchy have to be included. The same holds for a decrease of the lead temperature. It should be emphasized, though, that for a noninteracting system, i.e. $\lambda=0$, the hierarchy closes exactly at the second tier as long as only single-particle observables are considered. The approach is thus superior to basic perturbation theory, as demonstrated already in Ref.\ \onlinecite{Schinabeck2016}. 

For model 1 and $\Gamma=10^{-2} \unit{eV}$, the results obtained by a time-local truncation of the electronic hierarchy after the third (solid lines) and the second tier (dashed lines) agree very well (cf.\ Fig.\ \ref{fig:I_vib_V}). This shows that convergence is reached on the basis of the second tier of the electronic hierarchy. However, if molecule-lead coupling is increased by a factor of 10 (cf.\ Fig.\ \ref{fig:I_vib_V}c-f), there are deviations between the second and third-tier results for $\lambda/\Omega \geq 0.6$ and $\Phi < 1 \unit{eV}$. These deviations are most pronounced in the off-resonant transport regime which is governed by higher-order cotunneling processes.
As our current implementation of the approach VibBath only includes three tiers of the electronic hierarchy, we demonstrate that the third-level calculations represent the converged results by comparison to converged results of the approach VibSys provided in Fig.\ \ref{fig:I_vib_V_conv} in App.\ \ref{app:conv_vibsys}.

As already mentioned at the end of Sec.\ \ref{sec:theory_HEOM}, we apply a time-nonlocal closure in order to truncate the vibrational hierarchy. The number of vibrational tiers which has to be incorporated in order to guarantee convergence of the results in Fig.\ \ref{fig:I_vib_V} is summarized in Tab.\ \ref{tab:vib_tiers}. We can make a few general statements regarding the effort: \textit{i)} Higher electronic-vibrational coupling requires to include more tiers of the vibrational hierarchy.
\textit{ii)} In the resonant transport regime, a higher bias voltage, which results in more complex inelastic processes and a higher average vibrational excitation, also necessitates more vibrational tiers.
The average vibrational excitation as an observable is generally more difficult to convergence than the average current. This statement is demonstrated in App.\ \ref{app:conv_vib}.
\textit{iii)} Increasing the ratio $2 \Gamma/\Omega$ from the nonadiabatic to the crossover regime, the number of vibrational tiers is reduced for constant electronic-vibrational coupling $\lambda/\Omega$.
\textit{iv)} In the nonadiabatic and the crossover regime, we could converge results for $\lambda/\Omega \leq 1$, which required 30 tiers of the vibrational hierarchy, whereas in the adiabatic transport regime also stronger electronic-vibrational coupling up to $ \lambda/\Omega \approx 2$ can be treated. Due to the fast electron dynamics in the adiabatic limit, the influence of the slow vibration on the electron transport is effectively smaller than in the nonadiabatic regime. However, due to the higher vibrational excitation (cf. \textit{ii)}), more vibrational tiers have to be incorporated for $\Omega=0.05 \unit{eV}$ (adiabatic regime) than for $\Omega=0.2 \unit{eV}$ (crossover regime) at $\lambda/\Omega=1$.

%
Finally, the application range of the method VibBath is compared to that of the previously developed approach VibSys.\cite{Schinabeck2016} Within the approach VibSys, the vibrational mode is treated as part of the reduced system. Consequently, the resulting HQME have to be evaluated in an electronic-vibrational product basis.
The size of the vibrational basis set (e.g.\ eigenfunctions of the harmonic oscillator) determines the numerical effort.
For typical parameters, we could obtain converged results up to an average vibrational excitation of $\av{a^\dagger a} \lesssim 12$. This is reflected, e.g., in Fig.\ \ref{fig:I_vib_V_conv}, where converged results of the approach VibSys could sometimes not be achieved in the whole voltage range because of the large vibrational excitation. Thus, the approach VibSys cannot be applied to systems which exhibit high vibrational excitation, which is the case, e.g., in the limit of small frequency ($\Omega \ll \Phi$) or small electronic vibrational coupling ($\lambda/\Omega \to 0$). However, these systems with high vibrational excitation can be treated efficiently by our new approach VibBath as demonstrated before and in App.\ \ref{app:conv_prop}. The approach VibBath can also straightforwardly be extended to include multiple vibrational modes or a vibrational bath, which is not possible within the approach VibSys. On the other hand, the approach VibSys has the advantage that it is possible to describe strong electronic-vibrational coupling efficiently in the nonadiabatic transport regime. Additionally, the coupling to anharmonic vibrational modes can be treated, which is not possible within VibBath because the anharmonic environment cannot be integrated out analytically. Furthermore, if one is not only interested in the steady state but also in the transient dynamics, the method VibSys provides more flexibility in the choice of the initial state because the initial distribution of the vibrational excitation can be arbitrarily chosen whereas the approach VibBath is restricted to a thermal state for the vibration where only the temperature $T_\vib$ is arbitrary. 
Thus, depending on the model considered and the specific parameter regime, approach VibBath or VibSys may be more appropriate. The availability of both methods provides converged results in a broad range of parameters.
\clearpage
\newpage
\subsection{Vibrational instability in the regime of weak electronic-vibrational coupling} \label{sec:vib_instab}
In the second part of the results section, we apply the new methodology to study specifically resonant transport in the regime of weak electronic-vibrational coupling. This regime is characterized by the counter-intuitive phenomenon that the vibrational excitation can increase with decreasing electronic-vibrational coupling $\lambda$. In the limit $\lambda \to 0$ and for sufficiently high bias voltages, the vibrational excitation can assume very high values,\cite{Koch2006a,Haertle2011c,Kast2011,Haertle2015a} which may lead to the destruction of the nanomechanical system if the vibration is treated beyond the harmonic approximation.\cite{Gelbwaser2017} Therefore, this phenomenon is also referred to as vibrational instability.

While the phenomenon of the vibrational instability has been studied before,\cite{Koch2006a,Haertle2011c,Kast2011,Haertle2015a,Gelbwaser2017} the HQME methodology introduced above allows us, for the first time, to analyze it systematically with a numerically exact method, which includes in particular the influence of broadening induced by molecule-lead coupling on the distribution of the vibrational excitation for $\lambda / \Omega <1$ and in the limit $\lambda \to 0$. 
For analysis, 
the numerically exact HQME results are compared with analytical results of the Born-Markov master equation (BMME) for $\lambda \to 0$.\cite{Koch2006a,Gelbwaser2017} 
These results are based on a lowest-order expansion in $\Gamma$ and, thus, broadening of the electronic level due to molecule-lead coupling is neglected.
To characterize the distribution of vibrational excitation, we first study the dependence of the average excitation on electronic-vibrational, molecule-lead coupling and temperature in Sec.\ \ref{sec:av_exc}. This allows to distinguish between the influence of lead temperature and molecule-lead coupling. In Sec.\ \ref{sec:st_dev}, this study is complemented by the analysis of the second cumulant which corresponds to the width of the vibrational distribution. 
As a representative model system, the parameters of Ref.\ \onlinecite{Haertle2015a} are adopted and referred to as model 3 (cf.\ Tab.\ \ref{tab:models}), i.e. $\tilde \epsilon_0 = \epsilon_0 - \lambda^2 / \Omega=0.3 \unit{eV}$ is chosen for the reorganized energy level and $\Omega =0.1 \unit{eV}$ for the frequency of the vibrational mode at a bias voltage of $\Phi = 0.9 \unit{V}$.

\subsubsection{Average vibrational excitation} \label{sec:av_exc}

\begin{figure}[htbp]
	\centering
	\includegraphics[width=0.6\columnwidth]{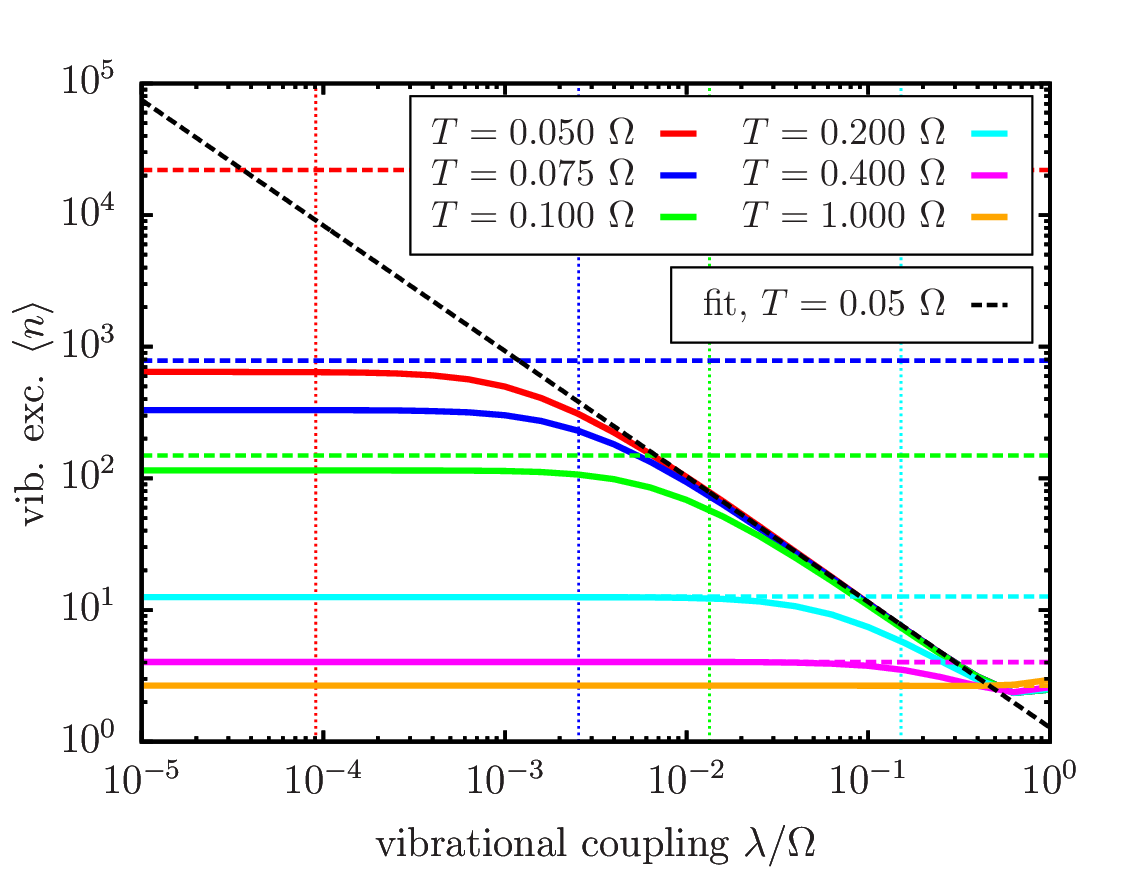}
	\caption{Average vibrational excitation $\av{n}$ as a function of electronic-vibrational coupling $\lambda/\Omega$ for different temperatures $T$ (solid lines). The results are obtained for model 3 and $\Gamma=10^{-2} \Omega$ on the basis of a truncation of the electronic hierarchy after the third tier with time-local truncation. The dashed horizontal and dotted vertical lines depict $\av{n}^\text{BMME}_{\lambda \to 0}$ and $\lambda^\text{BMME}_\text{thres} (T)$ as defined in Eqs.\ (\ref{eq:N_exp}) and (\ref{eq:lambda_thres}), respectively.  The colors indicate the corresponding temperatures. The dashed black curve represents a linear fit to the solid red line in the log-log plot.}
	\label{fig:N_lambda}
\end{figure}

We first study the average vibrational excitation $\av{n}$ as a function of electronic-vibrational coupling $\lambda/\Omega$ for different temperatures $T\equiv T_\leads$ of the leads and a small molecule-lead coupling $\Gamma=10^{-2} \Omega$. The respective results are depicted in Fig.\ \ref{fig:N_lambda}.
For $\lambda/\Omega \lesssim 0.5$ and temperatures $T \leq 0.4\, \Omega$, the vibrational excitation increases with decreasing electronic-vibrational coupling, which is a characteristic feature of a vibrational instability. Especially for low temperatures ($T \lesssim 0.1 \,\Omega$), the increase appears linear in the log-log plot, which indicates a power law behavior. For coupling strength $\lambda < \lambda_\text{thres} (T,\Gamma)$, where $\lambda_\text{thres} (T,\Gamma)$ denotes a threshold value, the vibrational excitation saturates and assumes a constant value $\av{n}_{\lambda \to 0}$ for $\lambda \to 0$. This value is strongly reduced with increasing temperature whereas $\lambda_\text{thres} (T,\Gamma)$ exhibits the opposite behavior.

In order to explain this behavior, we refer to analytic results which were previously derived within a perturbative BMME-treatment, which is based on a lowest order expansion in molecule-lead coupling $\Gamma$. Consequently, broadening effects due to molecule-lead coupling are neglected.
Assuming zero temperature and the strict wide-band limit ($\Gamma_K(\omega)=\text{const.}$), Koch \emph{et al.}\cite{Koch2006a} derived a scaling law for the vibrational excitation with respect to electronic-vibrational coupling $\lambda/\Omega$,  which is based on the following considerations:
For $\lambda / \Omega \ll 1$, 
processes, which change the vibrational excitation of the molecule, are less probable, the more vibrational quanta are (de)excited. As a result, the lowest-order inelastic processes ($\mathcal{O}(\lambda^2/\Omega^2)$) comprising the (de)excitation of one vibrational quantum $ n \to n \pm 1$ dominate (example processes are depicted in Fig.\ \ref{fig:sample_processes}a,b).
For our model parameters, the inelastic transport and pair creation processes $ n \to n \pm 1$ are possible with respect to the right lead whereas with respect to the left lead only the transport processes are enabled and the pair creation processes are blocked. Consequently, each excitation process is partnered by a deexcitation process which leads to a random walk through the ladder of vibrational states.\cite{Koch2006a} However, the next-to leading order inelastic processes ($\mathcal{O}(\lambda^4/\Omega^4)$) comprising two vibrational quanta break this symmetry: With respect to the left lead, the deexcitation of two vibrational quanta is possible in a transport process as well as in a pair creation process, whereas the corresponding excitation processes are blocked. The scaling law of Koch \emph{et al.} is based on this asymmetry.\cite{Koch2006a} For our parameter set, we obtain the following scaling behavior for the average vibrational excitation
  \begin{align}
 \av{n} \propto & \left( \frac{\lambda}{\Omega} \right)^{-b_1}
\label{eq:N_scal}
\end{align}
with $b_1=1$.
In order to check the correctness of this scaling law using the HQME results, the dashed black line in Fig.\ \ref{fig:N_lambda} represents a linear fit to the solid red line corresponding to $T=0.05 \Omega$ in the log-log plot. The fit is performed for the electronic-vibrational coupling range $\lambda / \Omega \in [10^{-2},10^{-1}]$.
The fit parameter is given by
\begin{align}
 b_1=& 0.953 \pm 0.06,
\end{align}
which is in very good agreement with the analytic prediction.
According to the scaling relation in Eq.\ (\ref{eq:N_scal}), the average vibrational excitation diverges with decreasing coupling $\lambda/\Omega \to 0$ for zero temperature. This is due to the fact that the next-to-leading order process comprising two vibrational quanta can be neglected in this limit so that the leading-order processes lead to the random walk behavior described above. However, for finite temperature $T \neq 0$, the creation of one electron-hole pair with respect to the left lead is enabled by the thermal broadening of the Fermi distribution (cf.\ Fig.\ \ref{fig:sample_processes}d).
The probability of this process ($\mathcal{O}(\lambda^2/\Omega^2)$) is higher than the probability for the next-to-leading order processes comprising two vibrational quanta for $\lambda < \lambda_\text{thres}^\text{BMME}(T)$  where $\lambda_\text{thres}^\text{BMME}(T)$ denotes the prediction for $\lambda_\text{thres}(T,\Gamma)$ within a BMME treatment. As a result, the scaling behavior breaks down for finite temperature and the average vibrational excitation becomes constant for $\lambda < \lambda_\text{thres}^\text{BMME}(T)$.

In the regime, where $\lambda/\Omega$ is small enough so that only leading order inelastic processes contribute, H\"artle and Kulkarni\cite{Haertle2015a} determined the vibrational distribution function analytically for our specific choice of model parameters, $\Gamma_\tL =\Gamma_\tR$ and $T < \Omega$. Recently, these findings were generalized by Gelbwaser-Klimovsky \emph{et al.}\cite{Gelbwaser2017} to arbitrary parameters. They showed that the vibrational distribution is given by the following geometric distribution
\begin{align}
 \rho_n =&  A^n \rho_0 = A^n (1-A)
 \label{eq:exp_dis}
\end{align}
with
\begin{align}
 A =& \frac{G^+(\epsilon_+) G^-(\epsilon_0) + G^-(\epsilon_-) G^+(\epsilon_0)}{G^+(\epsilon_-) G^-(\epsilon_0) + G^-(\epsilon_+) G^+(\epsilon_0)}
 \label{eq:A}
\end{align}
and
\begin{subequations}
\begin{align}
 G^\pm(\epsilon) =& \sum_K G^\pm_K(\epsilon),\\
 G^\pm_K(\epsilon) =& f (\pm(\epsilon-\mu_K)) \Gamma_K,\\
 \epsilon_\pm =& \epsilon_0 \pm \Omega.
\end{align}
\end{subequations}
%
%
Based on this distribution function, an expression for the average vibrational excitation can be derived
\begin{align}
 \av{n}^\text{BMME}_{\lambda \to 0}=& \sum_{n=0}^\infty n \rho_n= \frac{A}{1-A}.
 \label{eq:N_exp}
\end{align}
For our specific model parameters, Eq.\ (\ref{eq:N_exp}) simplifies to
\begin{align}
 \av{n}^\text{BMME}_{\lambda \to 0} =& \left(1 -f(\epsilon_0 + \Omega -\mu_\tL) \right)^{-1} + \frac{1}{2}
\label{eq:N_Haertle}
\end{align}
for $\Gamma_\tL =\Gamma_\tR$ as well as $T < \Omega$.
Eq.\ (\ref{eq:N_Haertle}) was originally derived by H\"artle and Kulkarni for this specific model system.\cite{Haertle2015a}

In this limit, we can also give an estimate of $\lambda^\text{BMME}_\text{thres}(T)$ introduced before within the BMME framework. To this end, we equate the probabilities of the competing next-to-leading order and thermally activated lowest order processes
\begin{align}
 \left(1 -f(\epsilon_0 + \Omega -\mu_\tL) \right) |X_{n \to n \pm 1}|^2 \approx |X_{n \to n \pm 2}|^2
\end{align}
where $|X_{n \to n \pm 1}|^2 \approx (n+1) \lambda^2$ and $|X_{n \to n \pm 2}|^2 \approx (n+1)(n+2) \lambda^4 /4$ are the Franck-Condon transition probabilities in the limit $n \lambda^2 \ll 1$.\cite{Koch2006}
Replacing $n$ by the average vibrational excitation $\av{n}^\text{BMME}_{\lambda \to 0}$ in Eq.\ (\ref{eq:N_Haertle}), we find
\begin{align}
 \lambda^\text{BMME}_\text{thres} (T) = 2 \left(1 -f(\epsilon_0 + \Omega -\mu_\tL) \right)
 \label{eq:lambda_thres}
\end{align}
for $\left(1 -f(\epsilon_0 + \Omega -\mu_\tL) \right) \ll 1$.
In Fig.\ \ref{fig:N_lambda}, $\av{n}^\text{BMME}_{\lambda \to 0}$ and $\lambda^\text{BMME}_\text{thres} (T)$ are depicted by dashed horizontal and dotted vertical lines, respectively.
As the probability for the relevant electron-hole pair creation process, which reduces the vibrational excitation, is proportional to $\left(1 -f(\epsilon_0 + \Omega -\mu_\tL) \right)$ and thus increases with temperature, $\av{n}^\text{BMME}_{\lambda \to 0}$ decreases with rising temperature according to Eq.\ (\ref{eq:N_Haertle}) whereas $\lambda^\text{BMME}_\text{thres} (T)$ exhibits the opposite behavior.

The values $\av{n}_{\lambda \to 0}$ obtained with the HQME method  agree with the analytic BMME prediction $\av{n}^\text{BMME}_{\lambda \to 0}$ only for $T \gtrsim 20 \Gamma$. For smaller temperatures, the HQME values are always lower where the largest difference is found for the lowest temperature. This finding demonstrates that the broadening of the electronic level due to molecule-lead coupling $\Gamma$, which is neglected within the BMME treatment, has a similar influence on the vibrational distribution as the thermal broadening of the Fermi distribution. It increases the probability for the electron-hole pair creation process depicted in Fig.\ \ref{fig:sample_processes}d.
This is also reflected by $\lambda_\text{thres} (T,\Gamma)$ which marks the transition between the plateau and the scaling region of the vibrational excitation. Similar to the limit values for $\lambda \to 0$, the Born-Markov prediction $\lambda^\text{BMME}_\text{thres}(T)$ is only valid for $T \gtrsim 10 \Gamma$; e.g.\ for $T=5\, \Gamma$ (red curve) the prediction is an order of magnitude too low.

\begin{figure}[htbp]
	\centering
\includegraphics[width=0.6\columnwidth]{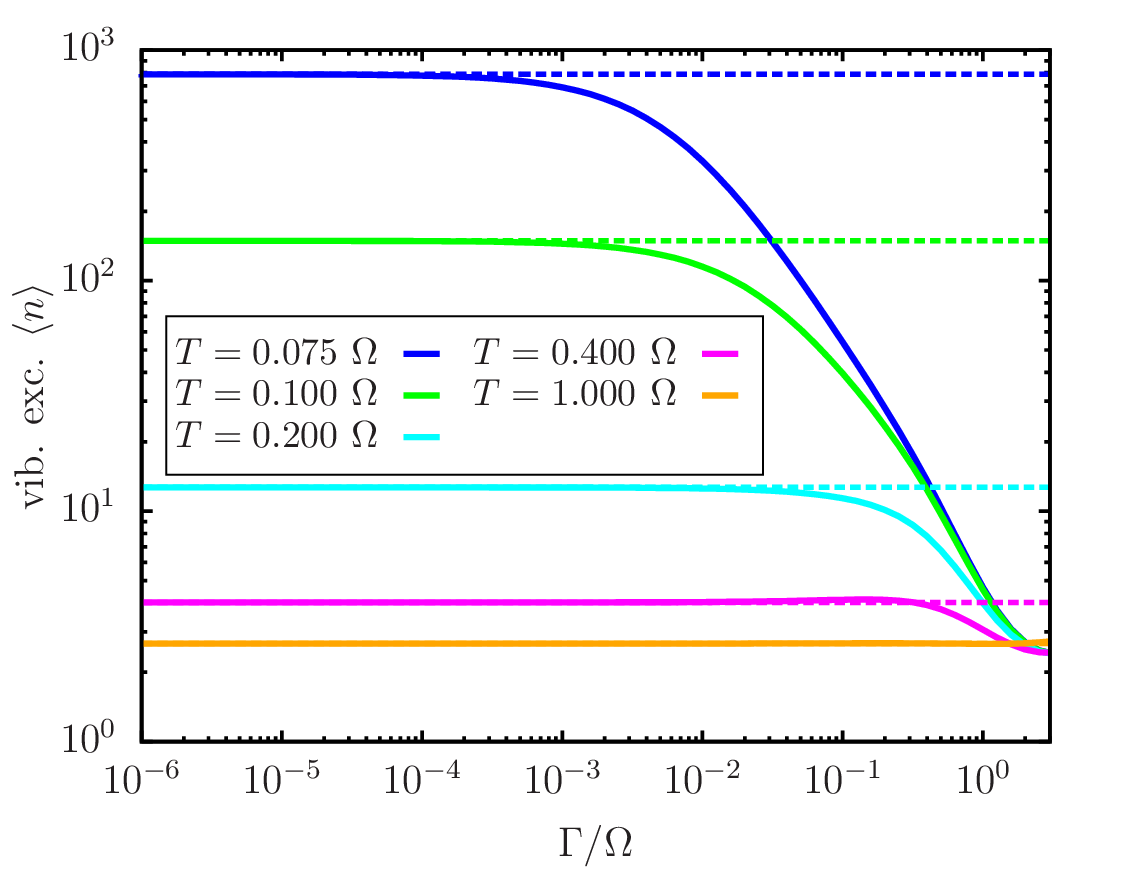}

  \caption{Average vibrational excitation $\av{n}$ as a function of $\Gamma$ for different temperatures depicted by solid lines. The results are obtained for model 3 and $\lambda/\Omega=10^{-5}$ on the basis of a truncation of the electronic hierarchy after the third tier with time-local truncation. The dashed horizontal lines represent the BMME limits $\av{n}^\text{BMME}_{\lambda \to 0}$ defined in Eq.\ (\ref{eq:N_exp}).}
  \label{fig:3rd_tier_gamma}
\end{figure}

In order to study the influence of molecule-lead coupling $\Gamma$ on the vibrational excitation in more detail in the limit $\lambda \to 0$, Fig.\ \ref{fig:3rd_tier_gamma} shows the average vibrational excitation as a function of $\Gamma$ for $\lambda/\Omega=10^{-5}$ and different temperatures $T$.
In the limit $\Gamma \to 0$ the accurate HQME results (solid lines) agree with the analytic result (dashed lines) of Eq.\ (\ref{eq:N_exp}), which was derived on the basis of a BMME.
If  $\Gamma$ is larger than a threshold value $\Gamma_{\text{thres}}$, which depends on temperature $T$, the vibrational excitation is no longer constant but decreases with increasing $\Gamma$. This confirms the above statement that an increased $\Gamma$ has a similar influence as an increased temperature $T$. However, the dependence of $\Gamma_{\text{thres}}$ on temperature $T$ is highly nonlinear: For $T/\Omega =0.075$ $(0.10)$, we find $\Gamma_{\text{thres}} / T \approx 10^{-3}$ $(3 \cdot 10^{-3})$, whereas $\Gamma_{\text{thres}}/T \approx 0.05$ holds for $T/\Omega =0.2$. This behavior demonstrates that the vibrational excitation is the more sensitive to finite molecule-lead coupling $\Gamma$, the lower the temperature is and thus the higher the limit value for $\Gamma \to 0$ is. The vibrational excitation for $T \leq 0.4 \Omega$ decreases with increasing $\Gamma$ until they reach a common value for $\Gamma / \Omega \approx 2$, which indicates that the influence of temperature can be neglected for $\Gamma \gtrsim 5 T$.  

\begin{figure}[htbp]
	\centering
\includegraphics[width=0.8\columnwidth]{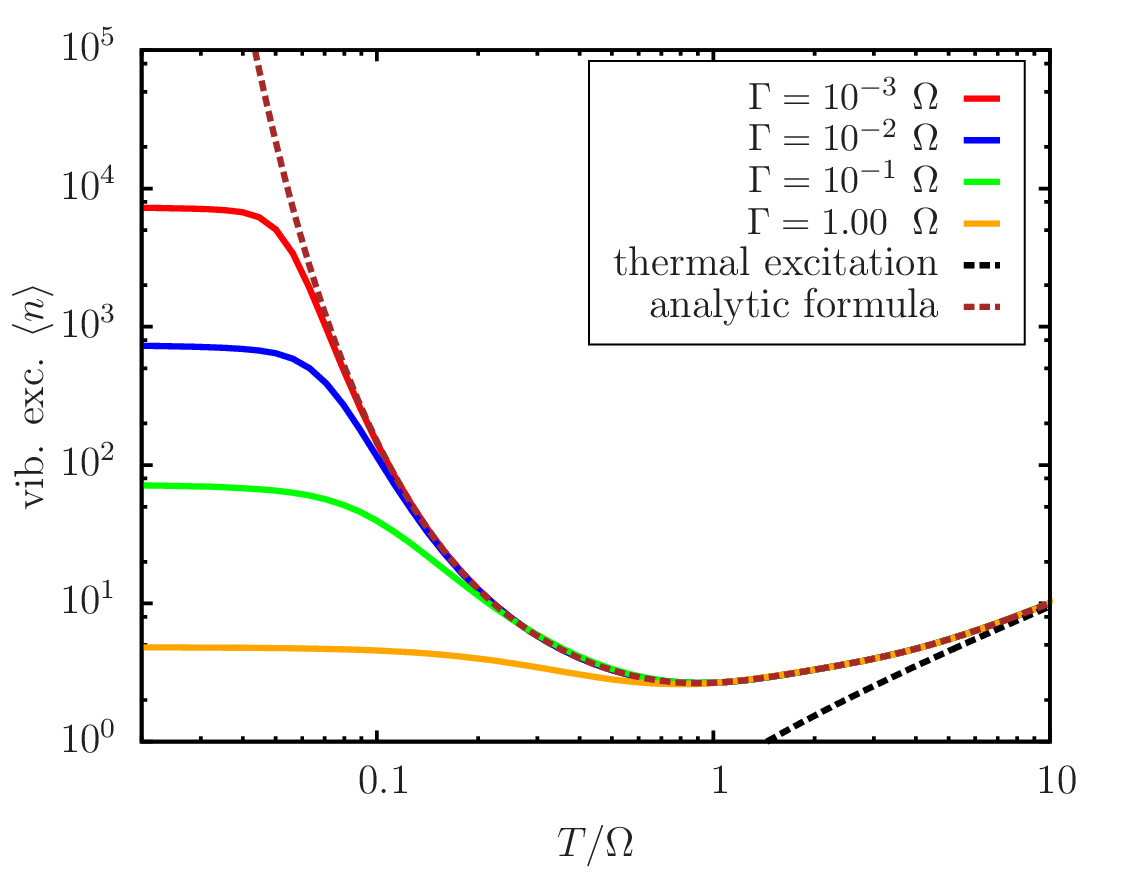}
  \caption{Average vibrational excitation $\av{n}$ as a function of temperature $T$ for different values of the molecule-lead coupling $\Gamma$. The electronic-vibrational coupling is chosen as $\lambda/\Omega=10^{-5}$. The solid lines correspond to a third tier truncation of the electronic hierarchy within the HQME-approach. The dashed brown and black curves depict the analytic BMME result of Eq.\ (\ref{eq:N_exp}) and the average thermal excitation, respectively.}
  \label{fig:3rd_tier_temp}
\end{figure}

%
%
After the systematic analysis  of the average vibrational excitation as a function of electronic-vibrational coupling $\lambda$ and molecule-lead coupling $\Gamma$, the dependence on the lead temperature $T$ is studied to complete the picture.
The respective HQME results are represented by solid lines in Fig.\ \ref{fig:3rd_tier_temp} for $\lambda/\Omega=10^{-5}$ and different values of $\Gamma$.
The dashed brown and black curves correspond to the analytic BMME result of Eq.\ (\ref{eq:N_exp}) and to the average thermal excitation, respectively. For $T/\Omega \gtrsim 10$, the latter agrees with the accurate HQME as well as the BMME result, which demonstrates that the vibrational excitation is solely determined by temperature.
For lower temperatures, $\av{n}$ follows the analytic predictions of Gelbwaser-Klimovsky in Eq.\ (\ref{eq:N_exp}) as long as $T > T_\text{thres}(\Gamma)$.
The threshold value $T_\text{thres}(\Gamma)$ is a strongly nonlinear function of molecule-lead coupling $\Gamma$, which reflects the discussion of $\Gamma_{\text{thres}}(T)$ in the last section: For low temperatures ($T /\Omega \lesssim 0.05$), already a relatively small molecule-lead coupling $\Gamma \gtrsim 10^{-2}\; T$ is sufficient to strongly influence the vibrational excitation.
For $T < T_\text{thres}(\Gamma)$, the average vibrational excitation assumes a constant value, which only depends on the molecule-lead coupling $\Gamma$. This suggests that in the limit $T \to 0$, a finite molecule-lead coupling $\Gamma$ leads to a finite vibrational excitation.

\begin{figure}[htbp]
	\centering
\includegraphics[width=0.8\columnwidth]{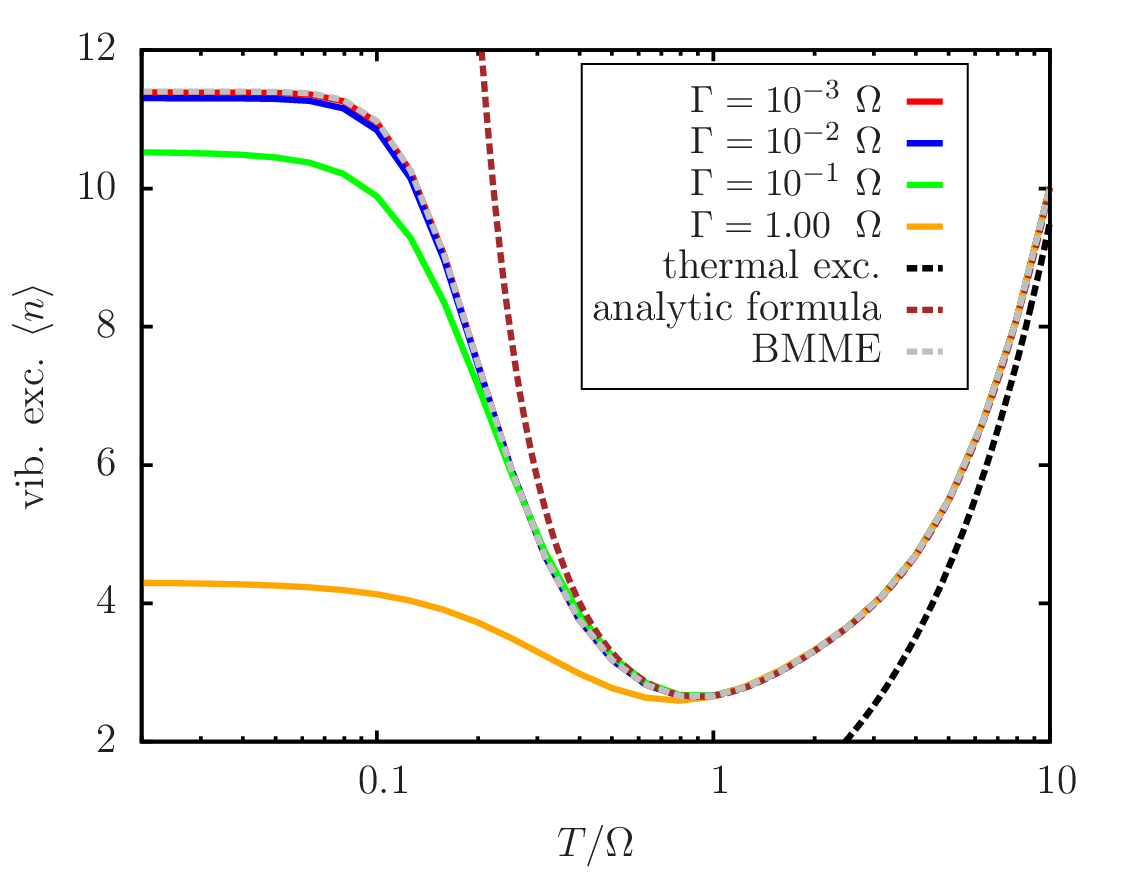}
  \caption{Average vibrational excitation $\av{n}$ as a function of temperature $T$ for different values of the molecule-lead coupling $\Gamma$. The electronic-vibrational coupling is chosen as $\lambda/\Omega=0.1$.The solid lines correspond to a third tier truncation of the electronic hierarchy within the HQME-approach. The dashed brown, gray and black curves depict the analytic BMME result of Eq.\ (\ref{eq:N_exp}), a full BMME calculation and the average thermal excitation, respectively.}
  \label{fig:3rd_tier_temp_01lo}
\end{figure}

In Fig.\ \ref{fig:3rd_tier_temp_01lo}, a higher electronic-vibrational coupling of $\lambda/\Omega=0.1$ is considered, which reveals a few interesting differences.
First of all, the full BMME calculation (gray dashed line) exhibits differences from the analytic formula (brown dashed line) for $T<\Omega$. This demonstrates that next-to leading order processes $\mathcal{O}\left( \lambda^4/\Omega^4 \right)$ have to be taken into account due to the increased electronic-vibrational coupling. The full BMME calculation saturates for $T< 0.1\, \Omega$ and assumes a constant value which shows that small lead temperatures do not influence the average vibrational excitation. This suggests that the vibrational excitation stays finite in the limit $T \to 0$ and $\Gamma \to 0$ for $\lambda/\Omega=0.1$, i.e.\ there is no random walk behavior through the ladder of vibrational states.
The numerically exact HQME results agree with the full BMME calculation for $\Gamma \lesssim 10^{-2}\, \Omega$, i.e. small molecule lead coupling - like small temperature - does not affect the vibrational excitation. Compared to the results for small coupling, $\lambda/\Omega=10^{-5}$ (Fig.\ \ref{fig:3rd_tier_temp} ), the level of vibrational excitation is reduced by almost three orders of magnitude in the limit $T \to 0$. For $\Gamma \gtrsim 10^{-1} \Omega$, this limit value decreases with increasing $\Gamma$.

\subsubsection{Vibrational distribution function} \label{sec:st_dev}
\begin{figure}[htbp]
	\centering
\begin{tabular}{ll}
a) & \\
&\includegraphics[width=0.6\columnwidth]{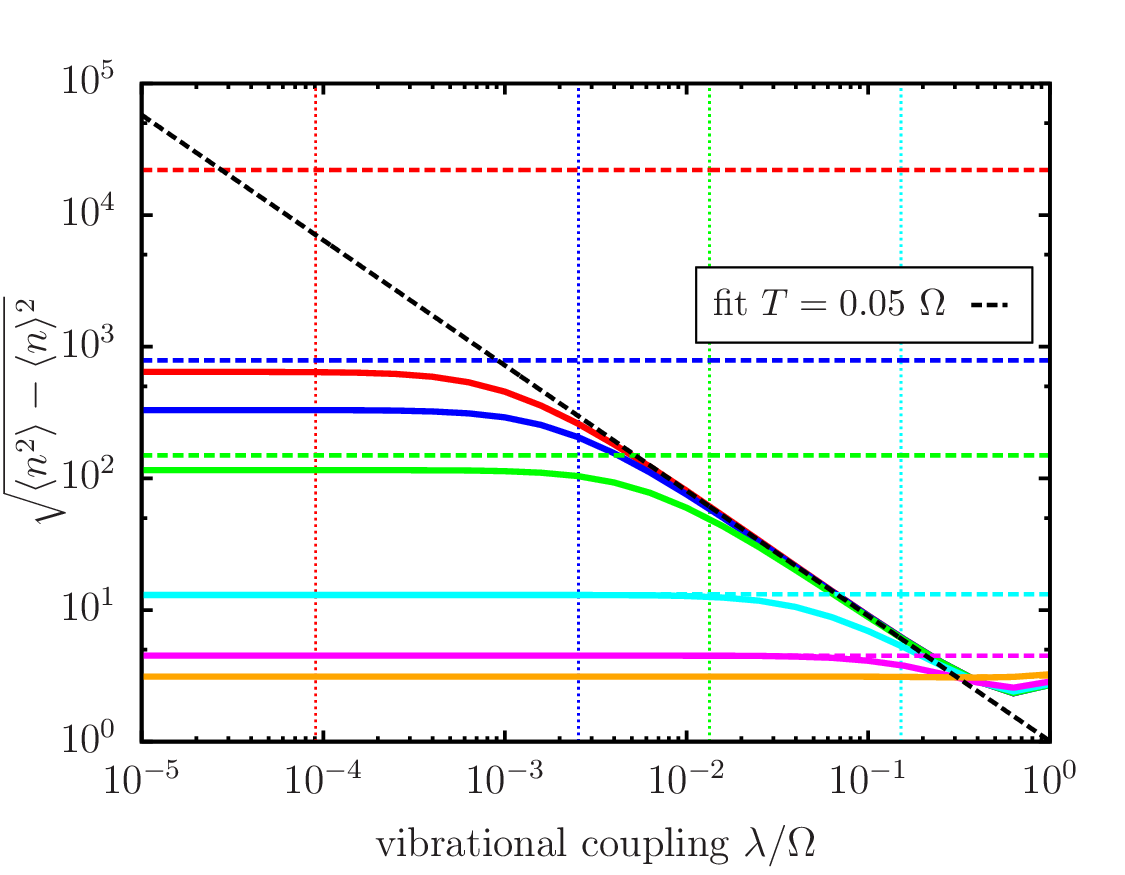}\\
b) & \\
&\includegraphics[width=0.6\columnwidth]{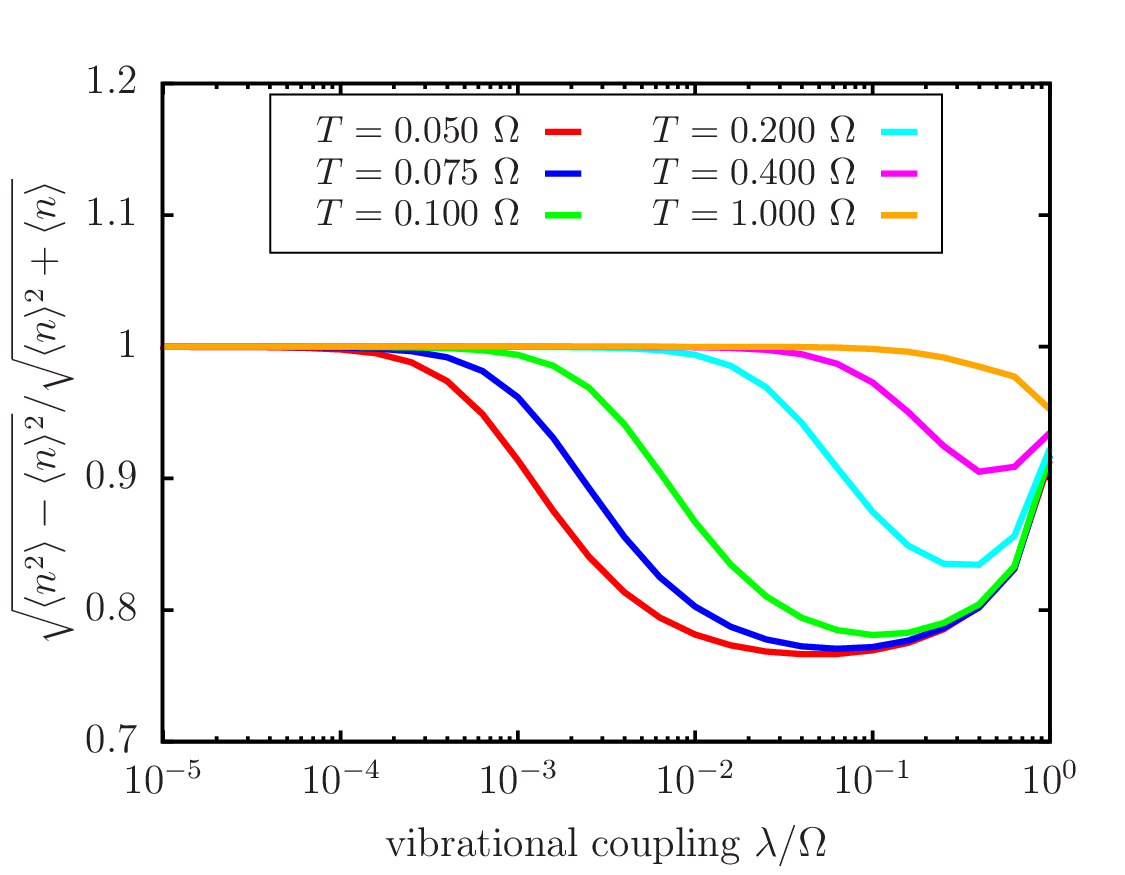}
\end{tabular}
\caption{Width $\sqrt{\av{n^2} -\av{n}^2}$ of the vibrational distribution (a) and the ratio $\sqrt{\av{n^2} -\av{n}^2} / \sqrt{\av{n}^2 + \av{n}}$ (b) are shown as a function of electronic-vibrational coupling $\lambda/\Omega$. The results are obtained for model 3 and $\Gamma=10^{-2} \Omega$ on the basis of a truncation of the electronic hierarchy after the third tier with time-local truncation.  In panel a), the dashed horizontal and dotted vertical lines depict $\sqrt{ \av{n^2} -\av{n}^2 } \big|^\text{BMME}_{\lambda \to 0}$ and $\lambda^\text{BMME}_\text{thres} (T)$ as defined in Eqs.\ (\ref{eq:Var_exp}) and (\ref{eq:lambda_thres}), respectively.  The colors indicate the corresponding temperatures. The dashed black curve represents a linear fit to the solid red line in the log-log plot.}
\label{fig:VarF_lambda}
\end{figure}

In order to gain more insight into the distribution function of the vibrational excitation beyond the average, the width of the distribution given by the standard deviation $\sqrt{\av{n^2} -\av{n}^2}$ is analyzed in the following. It is depicted in Fig.\ \ref{fig:VarF_lambda}a as a function of electronic-vibrational coupling for different temperatures.
The behavior of the width is mostly analogue to the average of the distribution $\av{n}$. The width is constant for $\lambda < \lambda_\text{thres}(T,\Gamma)$ and decreases for $\lambda_\text{thres}(T,\Gamma) < \lambda < 0.5\ \Omega$ following a power law.
According to the BMME predictions of Koch \emph{et al.},\cite{Koch2006a} it obeys the scaling relation
 \begin{align}
 \sqrt{\av{n^2} -\av{n}^2} \propto & \left( \frac{\lambda}{\Omega} \right)^{-b_2}
\label{eq:Var_scal}
\end{align}
with $b_2=1$. This conjecture is confirmed by the black dashed line in Fig.\ \ref{fig:VarF_lambda}a, which was obtained by a linear fit to the red curve ($T=0.05 \Omega$) in the log-log plot and gives the value $b_2= 0.951 \pm 0.002$.
In the limit $\lambda \to 0$, the width of the vibrational distribution in Eq.\ (\ref{eq:exp_dis}) is given by
\begin{align}
 \sqrt{ \av{n^2} -\av{n}^2 } \big|^\text{BMME}_{\lambda \to 0}=& \frac{\sqrt{A}}{(1-A)}=\sqrt{ \av{n}^2 + \av{n} } \big|^\text{BMME}_{\lambda \to 0},
 \label{eq:Var_exp}
\end{align}
where the last equality is universal for a geometric distribution. This expression reduces to
\begin{align}
 \sqrt{ \av{n^2} -\av{n}^2 } \big|^\text{BMME}_{\lambda \to 0}=& \sqrt{ \left(1 -f(\epsilon_0 + \Omega -\mu_\tL) \right)^{-2} - \frac{1}{4} }
\end{align}
for our specific model parameters, $\Gamma_\tL =\Gamma_\tR$ and $T < \Omega$. 
In Fig.\ \ref{fig:VarF_lambda}a, the values for $\sqrt{ \av{n^2} -\av{n}^2 } \big|^\text{BMME}_{\lambda \to 0}$, which are represented by dashed horizontal lines, show similar deviations from the numerically exact HQME results as  discussed above for the average vibrational excitation. These results suggest that a BMME treatment is only justified if $T \gtrsim 20\, \Gamma$.

In order to obtain more information on the nature of the vibrational distribution function, Fig.\ \ref{fig:VarF_lambda}b presents the ratio $\sqrt{ \av{n^2} -\av{n}^2 } / \sqrt{ \av{n}^2 + \av{n} }$ as a function of dimensionless electronic-vibrational coupling $\lambda/\Omega$. This observable combines the information of Fig.\ \ref{fig:N_lambda} and Fig.\ \ref{fig:VarF_lambda}a. According to Eq.\ (\ref{eq:Var_exp}), it is equal to unity in case of a geometric distribution and greater (smaller) than unity for a distribution which is wider (narrower) than a geometric distribution with the same average value.
This ratio is predicted to be one for $\lambda \to 0$ within a Born-Markov treatment because the vibrational excitation follows a geometric distribution according to Eq.\ (\ref{eq:exp_dis}). Remarkably, the HQME results in Fig.\ \ref{fig:VarF_lambda}b, which take the molecule-lead coupling into account, also exhibit this behavior for $\lambda \to 0$. This strongly indicates that the distribution is always geometric for $\lambda \to 0$, independently of the ratio $\Gamma/T$. For a rigorous proof, however, all cumulants would have to be analyzed.
For  $\lambda \gtrsim \lambda_\text{thres} (T,\Gamma)$, where $\lambda_\text{thres} (T,\Gamma)$ marks the transition between the constant and the scaling behavior of the vibrational excitation and the corresponding width, $\sqrt{ \av{n^2} -\av{n}^2 } / \sqrt{ \av{n}^2 + \av{n} }$ decreases below unity for all temperatures. This indicates that the vibrational distribution deviates from a geometric distribution and becomes more narrow for higher $\lambda$.

To complete the picture in the limit $\lambda \to 0$, we have analyzed the ratio \linebreak $\sqrt{\av{n^2} -\av{n}^2} / \sqrt{ \av{n}^2 + \av{n} }$ as a function of $\Gamma$ in analogy to Fig.\ \ref{fig:3rd_tier_gamma} (data not shown). We found that the results are constant and equal to unity over the investigated $\Gamma$ range. This suggests that the vibrational distribution is always geometric for arbitrary values of $\Gamma$ in the limit $\lambda \to 0$ and thus confirms the conjecture made above.

\section{Conclusion} \label{sec:conclusion}
In this paper, we have introduced a novel HQME approach (VibBath) for a numerically exact treatment of vibrationally coupled charge transport. The approach was applied to a generic model system comprising a single electronic state coupled to two macroscopic leads as well as to a single vibrational mode. The method VibBath is based on a system-bath partitioning, where the leads as well as the vibrational mode are treated as part of the bath subspace. 
This is in contrast to the HQME approach (VibSys) which we proposed in Ref.\ \onlinecite{Schinabeck2016}, where only the fermionic leads form the bath and the vibrational mode was treated as part of the reduced system. In comparison to VibSys, the novel method provides the advantage that systems with a large nonequilibrium vibrational excitation can be treated efficiently whereas within the approach VibSys the size of the Hilbert space of the reduced system is determined by the vibrational basis set and thus by the nonequilibrium vibrational excitation. This benefit comes at the cost, that the HQMEs include two coupled hierarchies of equations taking into account the coupling to the leads and to the vibration, respectively. 
Although the vibration is integrated out as part of the bath subspace, nonequilibrium effects are fully taken into account. This is in contrast to the approximate HQME method of Jiang \emph{et al.},\cite{Jiang2012} where due to the polaron transformation employed treating the vibration and the leads in the bath subspace is equivalent to neglecting the transport-induced nonequilibrium excitation of the vibration. In the approach VibBath, properties of the vibration, such as the moments of the vibrational distribution function, are encoded in the ADOs of the bosonic hierarchy and can thus be accessed without any additional computational effort. We have derived explicit expressions for the average vibrational excitation and the corresponding variance.

First, we have demonstrated that the HQME approach VibBath can be applied in a broad parameter space ranging from the nonadiabatic to the adiabatic transport regime and including both resonant and off-resonant transport. The convergence behavior with respect to the truncation of the vibrational hierarchy has been studied: On the one hand, with increasing electronic-vibrational coupling, more tiers have to be included and thus the numerical effort increases. On the other hand, a smaller number of vibrational tiers is necessary for convergence in the adiabatic transport regime compared to the nonadiabatic regime. As a result, the approach covers the regime of strong electronic-vibrational coupling (up to $\lambda/\Omega=2$) in the adiabatic transport regime by including typically less than 30 tiers of the vibrational hierarchy.

Second, we have investigated the influence of finite molecule-lead coupling on the nonequilibrium vibrational excitation in the regime of small electronic-vibrational coupling ($\lambda/\Omega \ll 1$).
Our numerically exact results extend former studies\cite{Koch2006a,Haertle2011c,Haertle2015a,Gelbwaser2017} which were based on a BMME treatment where the broadening of the electronic level due to molecule-lead coupling is neglected.
These studies reported that the average vibrational excitation and the corresponding variance can become singular in the limit of $\lambda/\Omega \to 0$ and zero temperature,\cite{Koch2006a} if the dominating electron-hole pair creation process is blocked by bias voltage.\cite{Haertle2011c} At finite temperature, this process is still enabled by the thermal broadening of the Fermi distribution, which leads to finite excitation.\cite{Haertle2015a} In this contribution, we have shown that the broadening of the electronic level due to molecule-lead coupling has a similar effect on the vibrational excitation as temperature. It leads to a further reduction of the average vibrational excitation and the corresponding variance as long as the normalized molecule-lead coupling $\Gamma/T$ is higher than a certain threshold value, which is not constant but depends on temperature. In particular for low temperatures, the vibrational excitation is more sensitive to a small value of $\Gamma/T$ than for high temperatures.
Additionally, our analysis of the first two moments suggests that the vibrational excitation is always described by a geometric distribution in the limit of $\lambda/\Omega \to 0$, independently of molecule-lead coupling $\Gamma$ and temperature $T$. This is a generalization of the BMME result of H\"artle and Kulkarni\cite{Haertle2015a} as well as Gelbwaser-Klimovsky \emph{et al.}.\cite{Gelbwaser2017}

\begin{acknowledgments}
We thank A.\ Erpenbeck, D.\ Gelbwaser-Klimovsky, and U.\ Peskin  for fruitful and inspiring discussions. This work was supported by the German Research Foundation (DFG) via SFB 953 and a research grant as well as the German-Israeli Foundation
for Scientific Research and Development (GIF). 
\end{acknowledgments}

\appendix
\section{Parametrization of the lead correlation function}\label{app:lead_correlation}
In order to express the thermal equilibrium correlation function $C^\sigma_{K} (t)$ of the free leads by a sum of exponentials (cf.\ Eq.\ (\ref{eq:C_leads_exp})), Eq.\ (\ref{eq:C_FT}) is used. $\Gamma_K (\omega)$ is assumed as a single Lorentzian as detailed in Eq.\ (\ref{eq:spec_dens}) and the Fermi distribution is approximated by a sum-over-poles scheme, the Pade decomposition.\cite{Hu2010,Hu2011,Zheng2012} This results in
\begin{align}
	f(x) & \approx \frac{1}{2} -\sum_{l=1}^{l_\text{max}} \frac{2 \kappa_l\ (x/T_\leads)}{(x/T_\leads)^2 + \xi_l^2} \equiv f_\text{approx}(x),
\label{app:fermi_approx}
\end{align}
where the derivation of the parameters $\kappa_l$ and $\xi_l$ can be found in Ref.\ \onlinecite{Hu2011}.
Consequently, the Fourier transform in Eq.\ (\ref{eq:C_FT}) can be performed by the theorem of residues and thus the following expressions for the parameters $\eta_{K,l}$ and $\gamma_{K,\sigma,l}$ in Eq.\ (\ref{eq:C_leads_exp}) are obtained
\begin{subequations}
\begin{align}
 \eta_{K,0}=&\frac{\Gamma_K W_K}{2} f_\text{approx}(\ii W),\\
\gamma_{K,\sigma,0}=&W_K -\sigma \ii \mu_K,\\
\eta_{K,l}=&- \ii T_\leads \kappa_l \cdot \frac{\Gamma_K W_K^2}{-\xi_l^2 T_\leads^2 +W_K^2}, \\
\gamma_{K,\sigma,l}=&\xi_l T_\leads -\sigma \ii \mu_K.
\end{align}
\end{subequations}
\section{Derivation of the hermiticity relation} \label{sec:herm_rel_vibbath}
In this appendix, the hermiticity relation in Eq.\ (\ref{eq:herm_rel_elvib}) is derived.
To this end, it is important to recall that the ADO $\rho_{j_p \cdots j_1|s_q \cdots s_1}^{(p,q)}$ can be expressed by the auxiliary Liouville propagator $J_{j_p \cdots j_1|s_q \cdots s_1}^{(p,q)}$ via Eq.\ (\ref{eq:ADO_def}).
Evaluating this relation in a basis of fermionic coherent states leads to
\begin{align}
\begin{split}
 \braket{\Phi_f| \rho_{j_p \cdots j_1|s_q \cdots s_1}^{(p,q)}(t) |\Phi'_{f}}=&\int \dd \Phi^*_i \dd \Phi_i \e^{-\Phi^*_i \Phi_i} \int \dd \Phi_i'^* \dd \Phi'_{i} \e^{-\Phi'^*_i \Phi'_i}\\
 & \times J_{j_p \cdots j_1|s_q \cdots s_1}^{(p,q)}(\Phi_f, \Phi'_{f},t; \Phi_i, \Phi'_{i},0) \braket{ \Phi_i | \rho (0)| \Phi'_{i} }.
 \end{split}
\end{align}
Consequently, the adjoint ADO is given by
\begin{align}
\begin{split}
 \braket{\Phi_f| \rho_{j_p \cdots j_1|s_q \cdots s_1}^{(p,q),\dagger}(t) |\Phi'_{f}} =& \braket{\Phi'_f| \rho_{j_p \cdots j_1|s_q \cdots s_1}^{(p,q)}(t) |\Phi_{f}}^* \\
=&\int \dd \Phi^*_i \dd \Phi_i \e^{-\Phi^*_i \Phi_i} \int \dd \Phi_i'^* \dd \Phi'_{i} \e^{-\Phi'^*_i \Phi'_i}\\
& \times J_{j_p \cdots j_1|s_q \cdots s_1}^{(p,q),*}(\Phi'_f, \Phi_{f},t; \Phi'_i, \Phi_{i},0) \braket{ \Phi_i | \rho (0)| \Phi'_{i} }.
\end{split}
\label{eq:nADO_conj}
\end{align}
where $\rho^\dagger =\rho$ has been used.
Applying the definition of $J_{j_p \cdots j_1|s_q \cdots s_1}^{(p,q)}$ in analogy to Eq.\ (\ref{eq:Liou_prop}), we find
\begin{align}
\begin{split}
 J_{j_p \cdots j_1|s_q \cdots s_1}^{(p,q),*}(\Phi'_f, \Phi_{f},t; \Phi'_i, \Phi_{i},0)=&\int_{\gv{\Phi}(0)=\gv{\Phi}_{i}}^{\gv{\Phi}^* (t)=\gv{\Phi}^*_{f}} \DD [\gv{\Phi}^* (t), \gv{\Phi}(t)]
\int_{\gv{\Phi}'(0)=\gv{\Phi}'_{i}}^{\gv{\Phi}'^* (t)=\gv{\Phi}'^*_{f}} \DD [\gv{\Phi}'^* (t), \gv{\Phi}'(t)]\\
& \times \expo{\ii \tilde S_\tS[\Phi,t]} \FF_{j_p \cdots j_1|s_q \cdots s_1}^{(p,q),*}[\Phi', \Phi,t] \expo{-\ii \tilde S_\tS[\Phi',t]}.
\end{split}
\label{eq:nJ_conj}
\end{align}
with
\begin{align}
\begin{split}
 \FF_{j_p \cdots j_1|s_q \cdots s_1}^{(p,q),*}&[\Phi',\Phi]=\left( \BB_{j_p}[\Phi',\Phi] \cdots \BB_{j_1}[\Phi',\Phi] \right)^* \left( \BB^\vib_{s_q}[\Phi',\Phi] \cdots \BB^\vib_{s_1}[\Phi',\Phi] \right)^* \FF^*[\Phi',\Phi] \\
=&\BB^*_{j_1}[\Phi',\Phi] \cdots \BB^*_{j_p}[\Phi',\Phi] \BB^{\vib,*}_{s_q}[\Phi',\Phi] \cdots \BB^{\vib,*}_{s_1}[\Phi',\Phi] \FF^*[\Phi',\Phi].
\end{split}
\label{eq:nIF_conj}
\end{align}
Thereby, we used that the order of Grassmann variables is reversed during complex conjugation. 
The equality $\FF^* [\Phi',\Phi] = \FF [\Phi,\Phi']$ follows from the hermiticity of the reduced density operator $\rho^\dagger=\rho$. Based on the definitions in Eqs.\ (\ref{eq:AB}), the following relations can be obtained 
\begin{subequations}
\begin{align}
\BB^*_j [\Phi',\Phi] = \BB_{\bar j} [\Phi,\Phi'],\\
\BB^{\vib,*}_s [\Phi',\Phi] = \BB^\vib_{\bar s} [\Phi,\Phi'].
\end{align}
\end{subequations}
Substituting these relations into Eq.\ (\ref{eq:nIF_conj}), leads to the following expression
\begin{align}
\begin{split}
 \FF_{j_p \cdots j_1|s_q \cdots s_1}^{(p,q),*}[\Phi',\Phi] =& \BB_{\bar j_1}[\Phi,\Phi'] \cdots \BB_{\bar j_p}[\Phi,\Phi'] \BB^\vib_{\bar s_q}[\Phi,\Phi'] \cdots \BB^\vib_{\bar s_1}[\Phi,\Phi'] \FF[\Phi,\Phi']\\
 \equiv& \FF^{(n)}_{\bar j_1 \cdots \bar j_p|\bar s_q \cdots \bar s_1 } [\Phi,\Phi'].
\end{split}
\label{eq:nIF_conj2}
\end{align}
Consequently, via Eqs.\ (\ref{eq:nJ_conj}) and (\ref{eq:nADO_conj}) a hermiticity relation for the ADO $\rho_{j_p \cdots j_1|s_q \cdots s_1}^{(p,q)}(t)$ is established
 \begin{align}
 \rho^{(p,q),\dagger} _{j_p \cdots j_1|s_q \cdots s_1}(t)=\rho^{(p,q)} _{\bar j_1 \cdots \bar j_p|\bar s_q \cdots \bar s_1} (t),
\label{eq:app_herm_rel_el}
\end{align}
which is identical to Eq.\ (\ref{eq:herm_rel_elvib}) after the permutation of the indices $j_\alpha$.
\section{Time-local truncation of the hierarchy} \label{app:closing}
In the following, we detail how the Markovian approximation of the ADOs of the anchor tier is performed, where we mainly follow Ref.\ \onlinecite{Xu2005}.
%
First, $\rho^{(p,q)}_{j_p \cdots j_1|s_q \cdots s_1}$ is formally written in a time-local form. For the molecule-lead coupling introduced in Eq.\ (\ref{eq:mol_lead_coup}), it reads
\begin{align}
  \rho^{(p,q)}_{j_p \cdots j_1|s_q \cdots s_1}(t) =& -\ii \left(  \eta_{K_p,l_p} B_{j_p}(t) \rho^{(p-1,q)}_{j_{p-1} \cdots j_1|s_q \cdots s_1} (t) - (-)^p  \eta_{K_p,l_p}^* \rho^{(p-1,q)}_{j_{p-1} \cdots j_1|s_q \cdots s_1} (t) B_{j_p}(t) \right)
\label{eq:rho_timelocal}
\end{align}
with
\begin{align}
   B_j ( t ) =& \int_0^t \dd \tau \e^{-\gamma_j \tau} \e^{\ii ( H_\tS + H_\tSB)(-\tau) } d^\sigma \e^{-\ii ( H_\tS + H_\tSB)(-\tau) }.
\end{align}
As it is not possible to evaluate Eq.\ (\ref{eq:rho_timelocal}) directly, the Markovian approximation for $B_j(t)$ is used, which leads to
\begin{align}
   B_j^\infty =& \int_0^\infty \dd \tau \e^{-\gamma_j \tau} \e^{\ii H_\tS (-\tau) } d^\sigma \e^{-\ii H_\tS(-\tau) },
\end{align}
where $\e^{\pm \ii ( H_\tS + H_\tSB)(-\tau) } \approx \e^{\pm \ii H_\tS(-\tau) }$ is applied and the upper integration limit is set to infinity $t \to \infty$.
Focusing only on the steady state regime, the second step is not an approximation.
%
%
For the electronic hierarchy in VibBath the operator $B_j^\infty$ assumes the form
\begin{align*}
  B^\infty_j ( t )=&\int_0^\infty \dd \tau \e^{- (\gamma_j + \ii \sigma \epsilon_0)  \tau} d^\sigma = \frac{ 1 }{\gamma_j + \ii \sigma \epsilon_0} d^\sigma.
\end{align*}
\section{Convergence properties} \label{app:conv_prop}
In this appendix, the convergence properties of the approach VibBath are analyzed. To this end, the convergence of the results with respect to the number of exponential terms used in the parametrization of the two time-bath correlation function as well as the truncation of the vibrational hierarchy is demonstrated in Apps.\ \ref{app:conv_Pade} and \ref{app:conv_vib} on the basis of representative examples. In App.\ \ref{app:conv_vibsys}, the results presented in Fig.\ \ref{fig:I_vib_V} are compared to the approach VibSys in order to prove convergence.
\subsection{Convergence with respect to the number of Pade poles} \label{app:conv_Pade}
\begin{figure}[h!]
\begin{tabular}{cc}
\includegraphics[width=0.5\columnwidth]{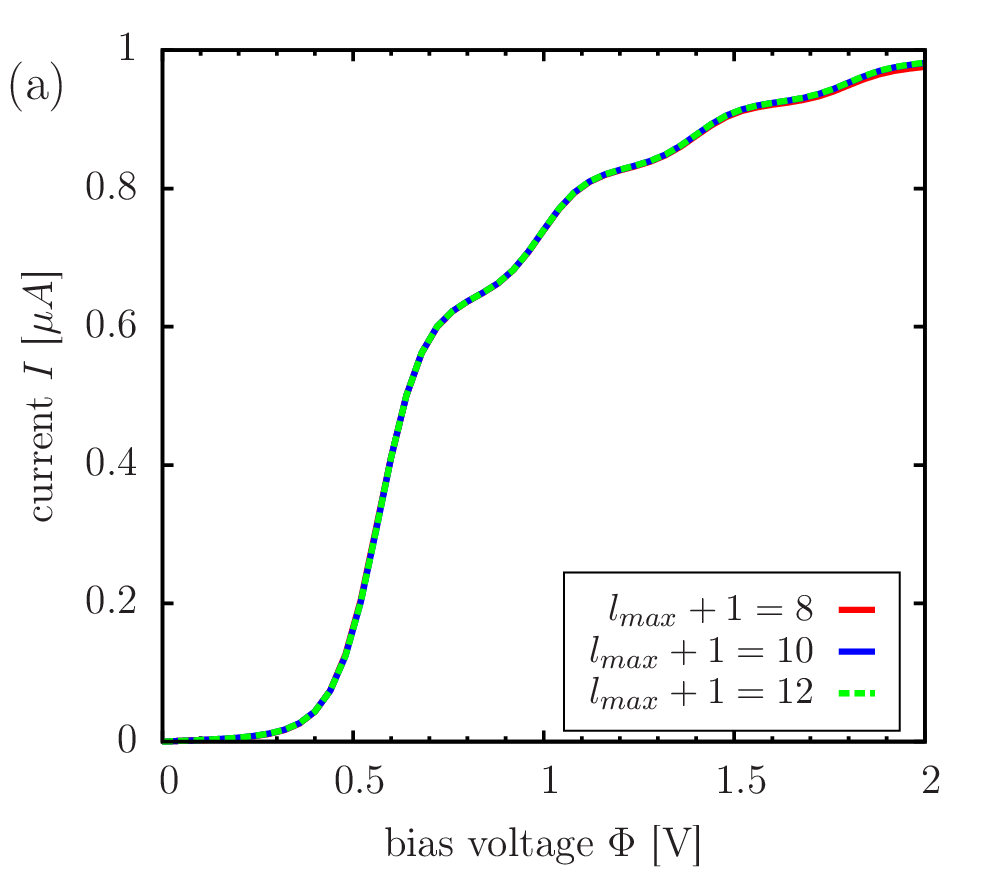}\\
\includegraphics[width=0.5\columnwidth]{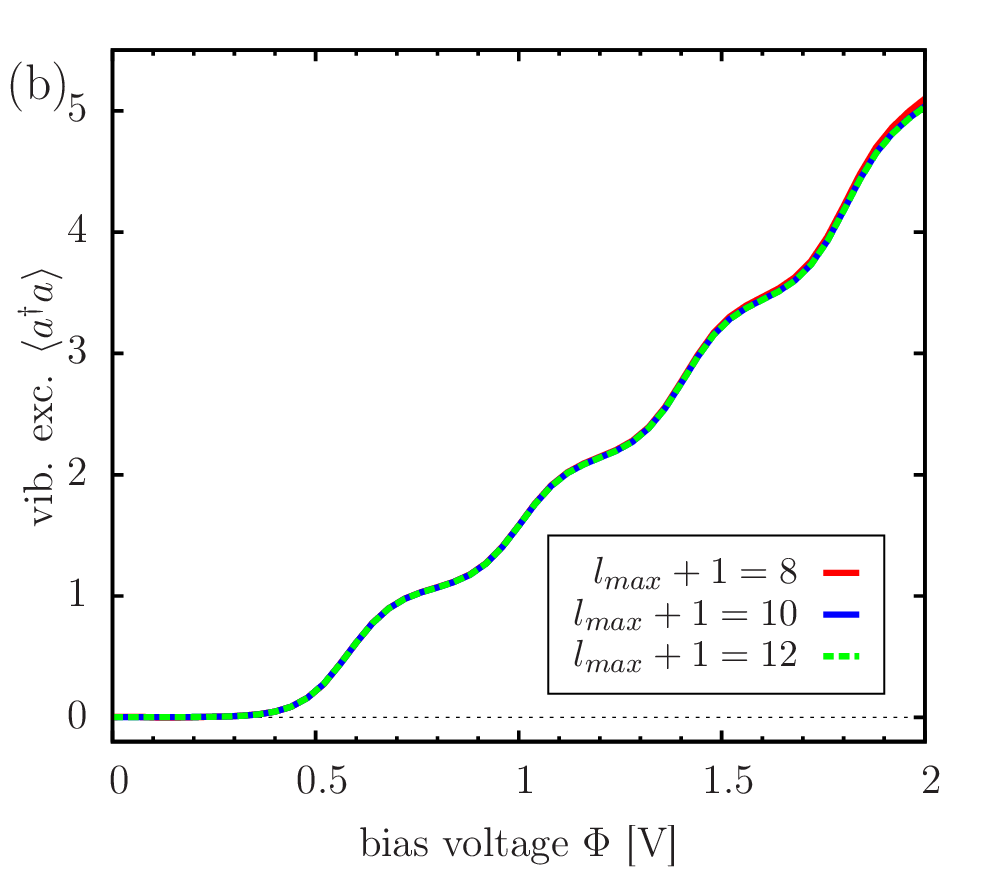}
\end{tabular}
\caption{Convergence of the current-voltage (a) and the vibrational excitation-voltage characteristics (b) for model 1, $\Gamma = 10^{-2} \unit{eV}$  and $\lambda/\Omega=1$ with respect to the number of exponential terms ($l_\text{max} + 1$) in the expansion of the lead correlation function  $C^\sigma_K (t)$.}
\label{fig:I_vib_Pade}
\end{figure}

In order to derive a closed set of HQME, the bath correlation function of the noninteracting leads is approximated by $(l_\text{max} + 1)$ exponential terms in Eq.\ (\ref{eq:C_leads_exp}). As outlined in App.\ \ref{app:lead_correlation}, the $(l=0)$-exponential term originates from the Lorentzian spectral density in Eq.\ (\ref{eq:spec_dens}) and the other $l_\text{max}$ terms stem from the Pade approximation for the Fermi distribution in Eq.\ (\ref{app:fermi_approx}).
Fig.\ \ref{fig:I_vib_Pade} illustrates the convergence of the observables of interest with respect to the number $(l_\text{max}+1)$ of exponential terms which are used for the approximation of the bath correlation function. The current as well as the vibrational excitation are depicted as a function of bias voltage for model 1, $\Gamma=10^{-2} \unit{eV}$ and $\lambda /\Omega =1$ (corresponding to the solid green line in Fig.\ \ref{fig:I_vib_V} a,b and $l_\text{max}$ is varied. Both observables are converged for $(l_\text{max}+1)=10$.
\subsection{Convergence with respect to the truncation of the vibrational hierarchy} \label{app:conv_vib}
\begin{figure}[h!]
\begin{tabular}{c}
\includegraphics[width=0.6\columnwidth]{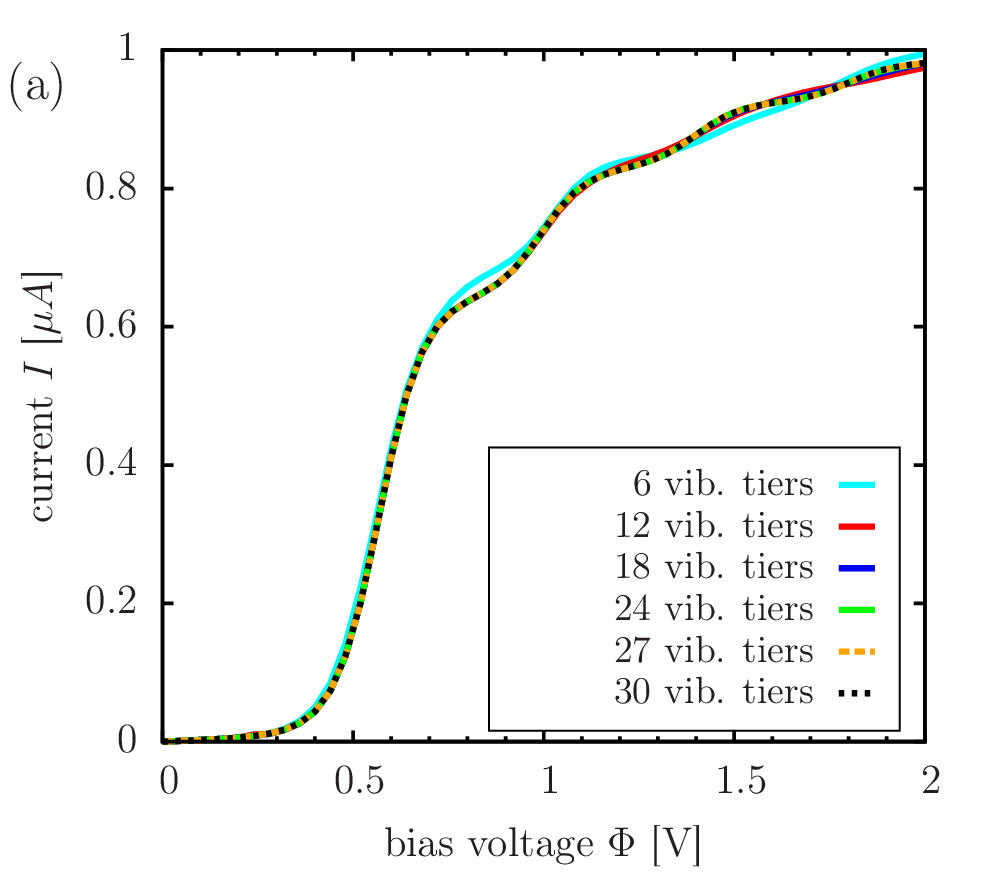}\\
\includegraphics[width=0.6\columnwidth]{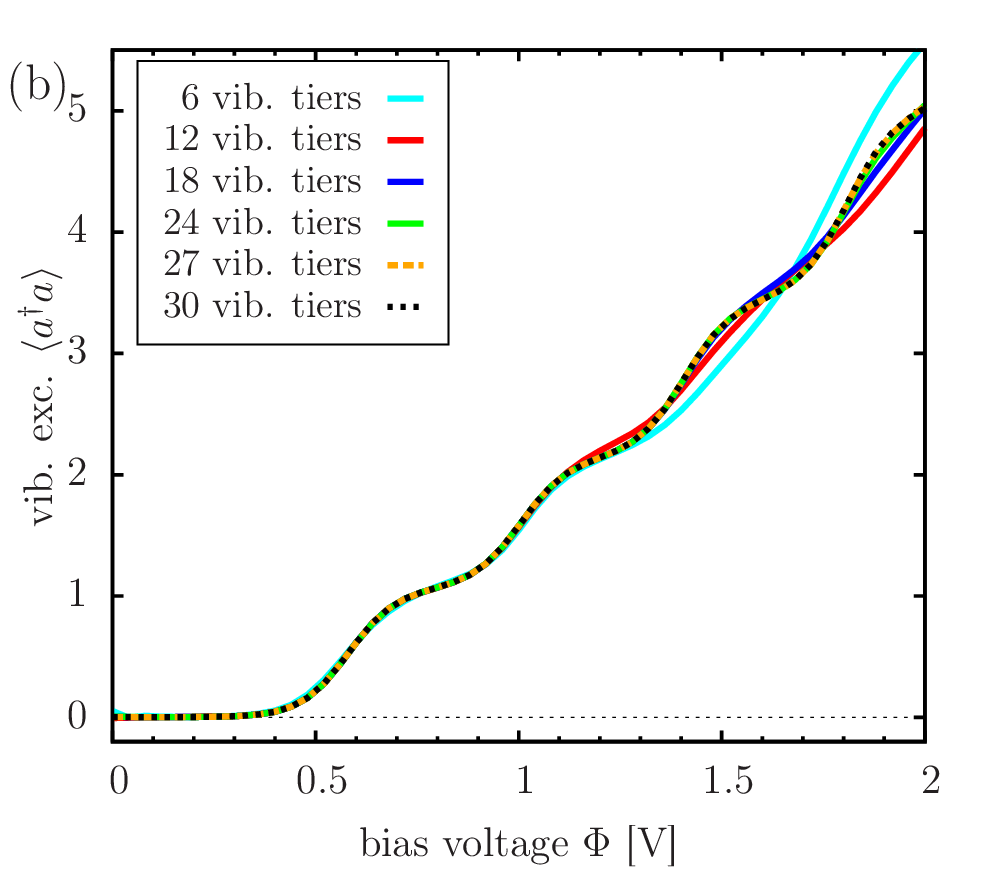}
\end{tabular}
\caption{Convergence of the current-voltage (a) and the vibrational excitation-voltage characteristics (b) for model 1, $\Gamma = 10^{-2} \unit{eV}$ and $\lambda/\Omega=1$ with respect to the truncation level of the vibrational hierarchy.}
\label{fig:I_vib_vibtiers}
\end{figure}

Fig.\ \ref{fig:I_vib_vibtiers} illustrates the convergence of the current-voltage and vibrational excitation-voltage characteristics with respect to the truncation of the vibrational hierachy on the basis of model 1, $\Gamma=10^{-2} \unit{eV}$ and $\lambda /\Omega =1$. As already mentioned in Sec.\ \ref{sec:applicability}, the current-voltage characteristics is easier to converge than the vibrational excitation-voltage characteristics. The current as a function of bias voltage is converged for a time-local truncation of the vibrational hierarchy after 27 tiers, whereas 30 tiers are necessary for the vibrational excitation.
The data also demonstrate that the higher the bias voltage and thus the vibrational excitation is, the more tiers of the vibrational hierarchy have to be incorporated to achieve convergence in the resonant transport regime.

%
%
\subsection{Convergence with respect to the electronic hierarchy} \label{app:conv_vibsys}
\begin{figure*}[h!]
\begin{tabular}{cc}
\includegraphics[width=0.45\textwidth]{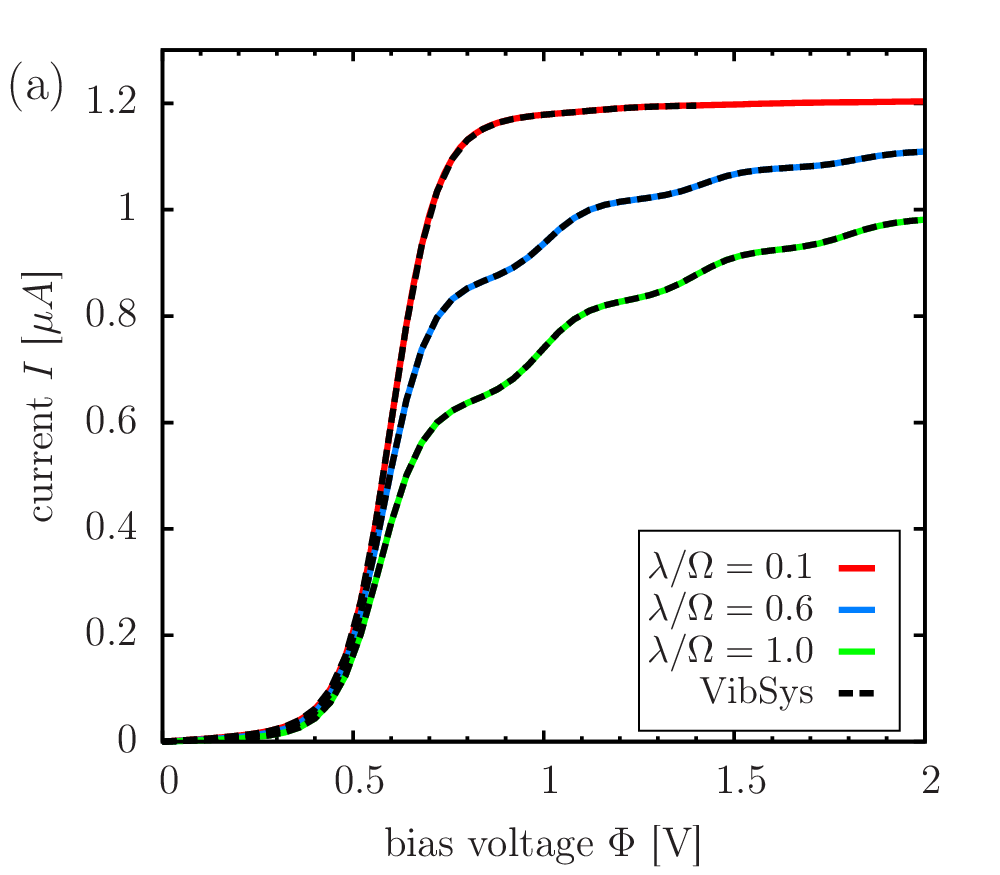}
\includegraphics[width=0.45\textwidth]{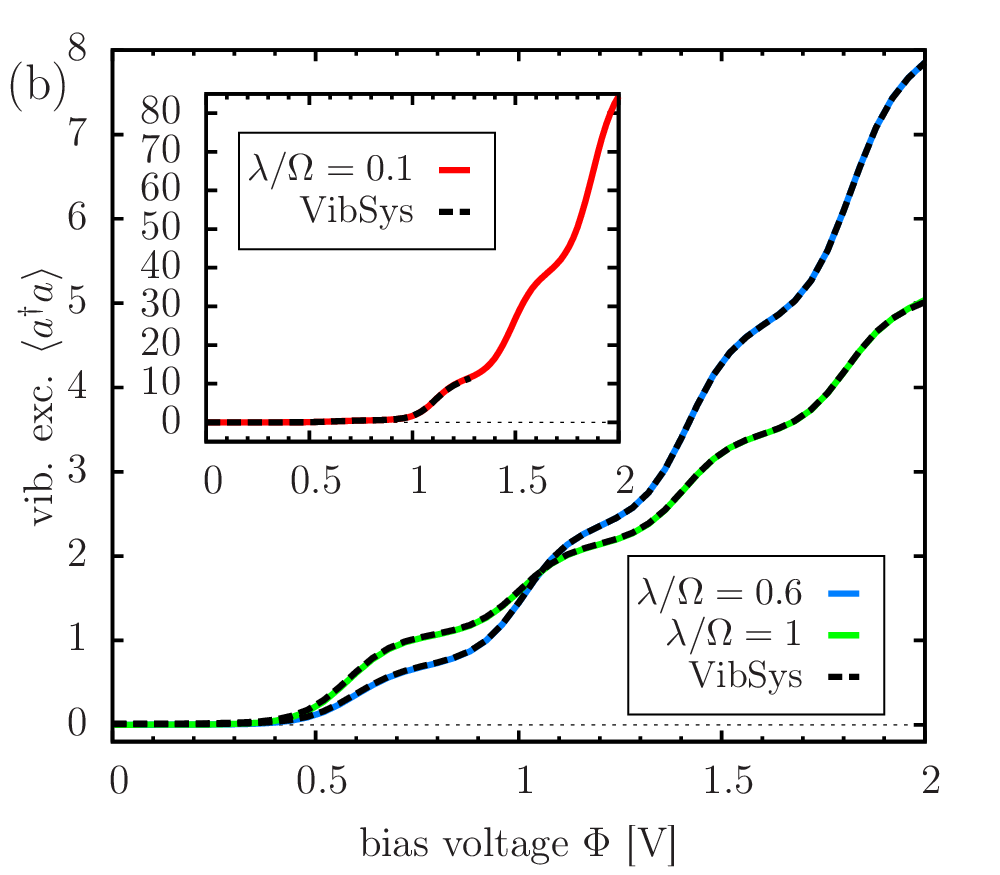}\\
%
\includegraphics[width=0.45\textwidth]{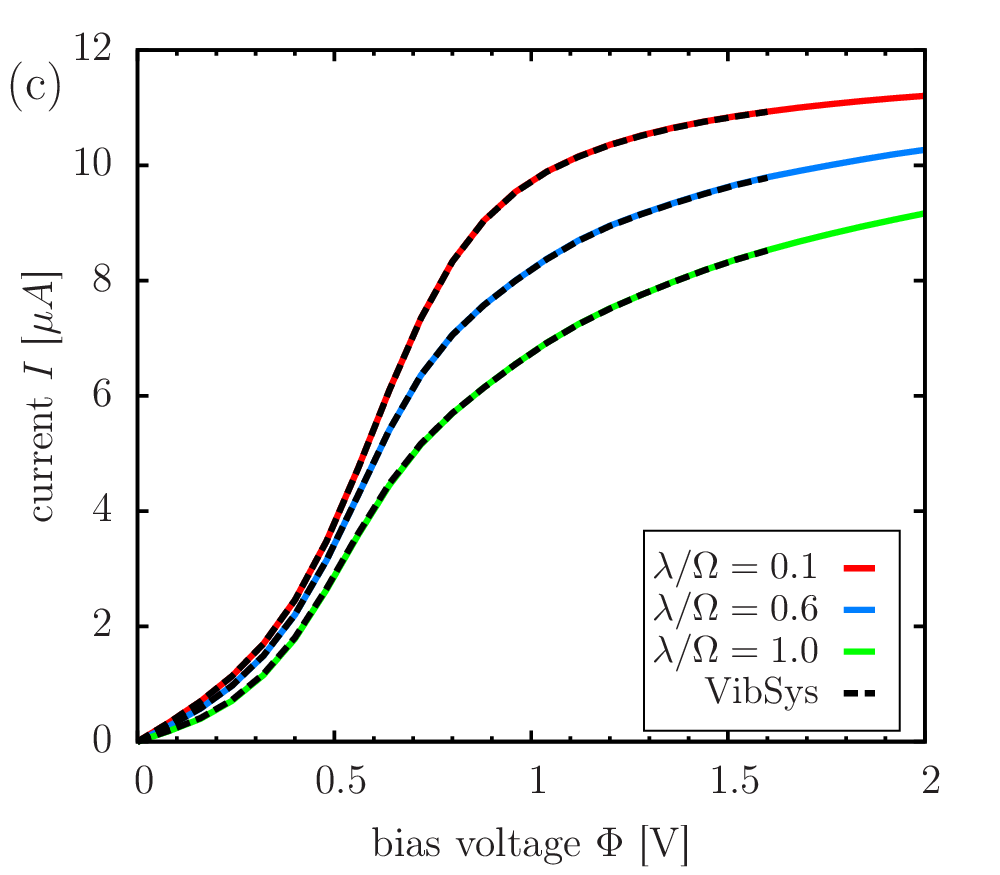}
\includegraphics[width=0.45\textwidth]{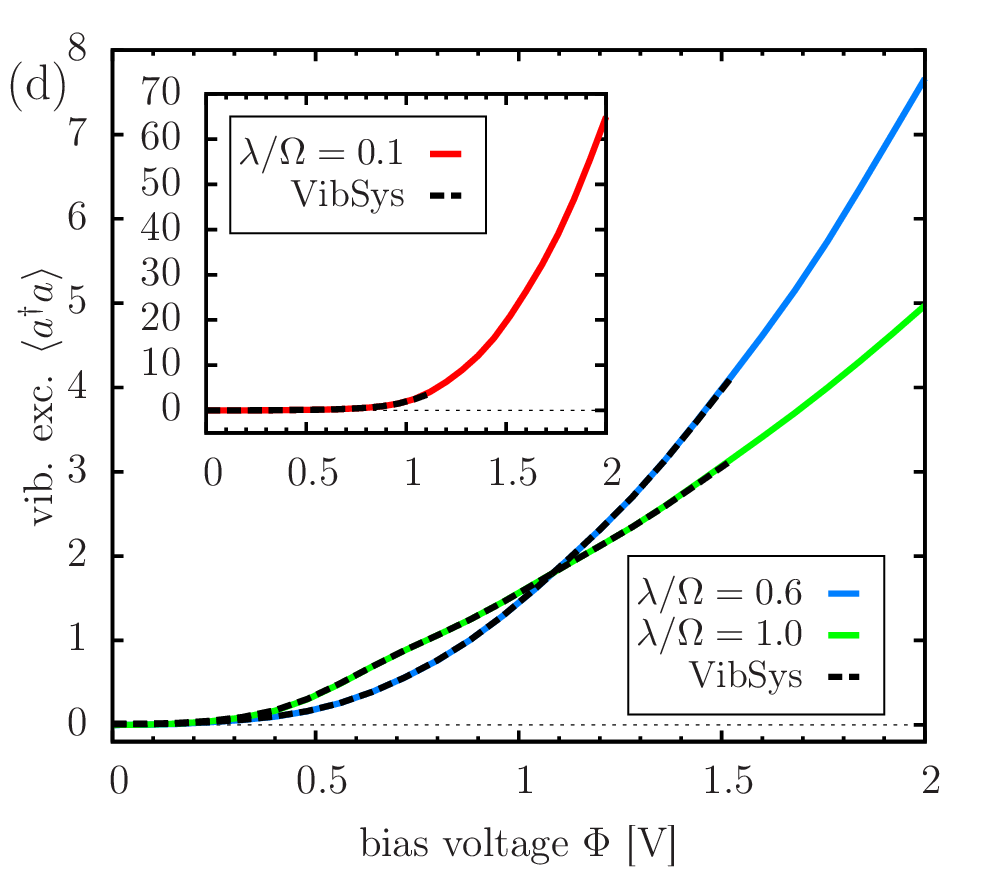}\\
%
\includegraphics[width=0.45\textwidth]{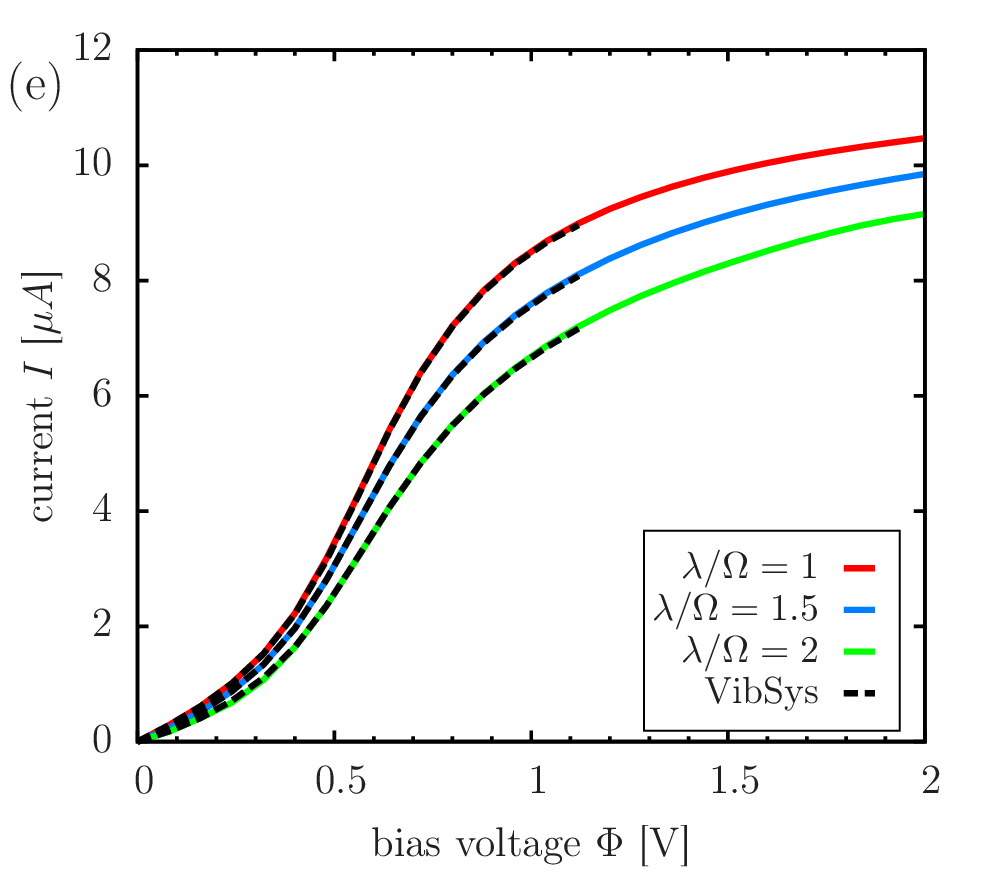}
\includegraphics[width=0.45\textwidth]{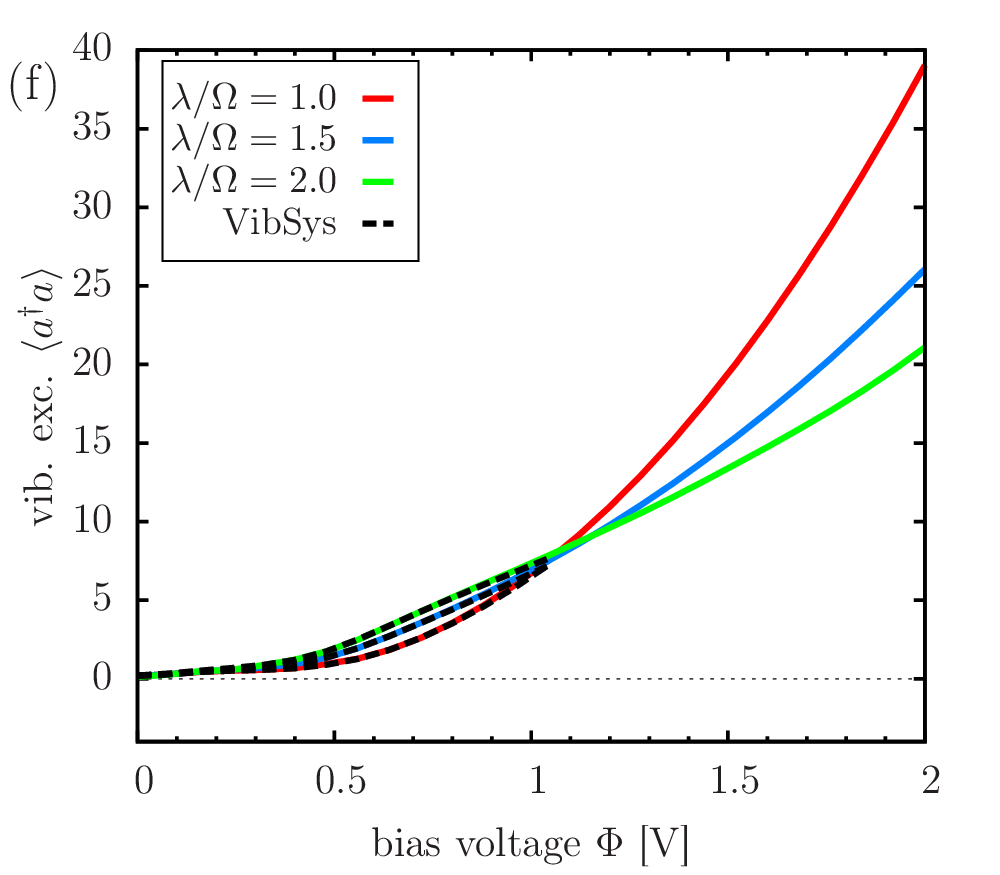}
\end{tabular}
\caption{Comparison of the results presented in Fig.\ \ref{fig:I_vib_V} to converged results of the approach VibSys (black dashed lines). The solid lines are obtained by the approach VibBath and a time-local truncation of the electronic hierarchy after the third tier. The current-voltage and vibrational excitation-voltage characteristics are shown for different electronic vibrational couplings $\lambda/\Omega$ and correspond to model 1 (cf.\ Tab.\ \ref{tab:models}) and $\Gamma=0.01 \unit{eV}$ (a,b) as well as $\Gamma=0.1 \unit{eV}$ (c,d), and model 2 and $\Gamma=0.1 \unit{eV}$ (e,f).}
\label{fig:I_vib_V_conv}
\end{figure*}

In the following, it is shown that the results presented in Fig.\ \ref{fig:I_vib_V} are converged with respect to the truncation of the electronic hierarchy. As we have implemented the approach VibBath only up to a time-local truncation of the electronic hierarchy at the third tier, the HQME approach VibSys introduced in Ref.\ \onlinecite{Schinabeck2016} is used to confirm convergence if the second and third tier truncations of VibBath do not agree. To this end, Fig.\ \ref{fig:I_vib_V_conv} shows the third tier results obtained by VibBath (solid lines) as well as the converged results of VibSys (black dashed lines).
Within the approach VibSys the vibration is treated as part of the reduced system so that the corresponding HQME have to be evaluated within a truncated basis set for the vibrational subspace. This prevents the use of this method in the regime of high vibrational excitation as already detailed in Sec.\ \ref{sec:applicability}. Consequently, the black dashed lines do often not cover the regime of high bias voltage (high vibrational excitation) in Fig.\ \ref{fig:I_vib_V_conv}. However, for high bias voltages $\Phi \gtrsim 1 \unit{V}$, the results corresponding to a time-local truncation of the electronic hierarchy after the second and third-tier agree very well. Consequently, the convergence of the third-tier results in Fig.\ \ref{fig:I_vib_V} is guaranteed in the whole voltage range.

%
%
%
%
\clearpage
\bibliography{./mybib}

\end{document}